\documentclass{aa}
\usepackage{natbib}
\usepackage[varg]{txfonts}
\usepackage{graphicx}

\usepackage{color} 

\def\Msun{\ifmmode{\mathrm M_\odot}\else{M$_\odot$}\fi}

\def\v2#1{\textcolor{blue}{\textbf{#1}}}
\def\rev#1{#1}

\begin{document}

\title{AGN feedback in the nucleus of M51}

\author{M.~Querejeta\inst{1}
\and E.~Schinnerer\inst{1}
\and S.~Garc\'{i}a-Burillo\inst{2}
\and F.~Bigiel\inst{3}
\and G.~A.~Blanc\inst{4,5,6}
\and D.~Colombo\inst{7}
\and A.~Hughes\inst{8,9}
\and K.~Kreckel\inst{1}
\and A.~K.~Leroy\inst{10}
\and S.~E.~Meidt\inst{1}
\and D.~S.~Meier\inst{11}
\and J.~Pety\inst{12,13}
\and K.~Sliwa\inst{1}
}

\institute{Max Planck Institute for Astronomy, K\"{o}nigstuhl, 17, 69117 Heidelberg, Germany,  \email{querejeta@mpia-hd.mpg.de}
\and Observatorio Astron\'{o}mico Nacional, Alfonso XII, 3, 28014 Madrid, Spain
\and Institut f\"{u}r theoretische Astrophysik, Zentrum f\"{u}r Astronomie der Universit\"{a}t Heidelberg, Albert-Ueberle-Str.~2, 69120 Heidelberg, Germany
\and Departamento de Astronom\'{i}a, Universidad de Chile, Camino del Observatorio 1515, Las Condes, Santiago, Chile
\and Centro de Astrof\'{i}sica y Tecnolog\'{i}as Afines (CATA), Camino del Observatorio 1515, Las Condes, Santiago, Chile
\and  Visiting Astronomer, Observatories of the Carnegie Institution for Science, 813 Santa Barbara St, Pasadena, CA 91101, USA
\and  Max Planck Institute for Radioastronomy, Auf dem H\"{u}gel 69, 53121 Bonn, Germany
\and CNRS, IRAP, 9 Av. Colonel Roche, BP 44346, 31028 Toulouse, France
\and Universit\'{e} de Toulouse, UPS-OMP, IRAP, 31028 Toulouse, France 
\and Department of Astronomy, The Ohio State University, 140 West 18th Avenue, Columbus, OH 43210, USA
\and Physics Department, \rev{New Mexico Institute of Mining and Technology}, 801 Leroy Place, Socorro, NM 87801, USA
\and Institut de Radioastronomie Millim\'{e}trique, 300 Rue de la Piscine, 38406 Saint Martin d'H\`{e}res, France
\and Observatoire de Paris, 61 Avenue de l’Observatoire, 75014 Paris, France
}

\date{Received ..... / Accepted .....}

\abstract {
AGN feedback is invoked as one of the most relevant mechanisms that shape the evolution of galaxies.
Our goal is to understand the interplay between AGN feedback and the interstellar medium in M51, a nearby spiral galaxy with a modest AGN and a kpc-scale radio jet expanding through the disc of the galaxy. For that purpose, we combine molecular gas observations in the \mbox{CO(1-0)} and \mbox{HCN(1-0)} lines from the Plateau de Bure interferometer with archival radio, X-ray, and optical data.
We show that there is a significant scarcity of CO emission in the ionisation cone, while molecular gas emission tends to accumulate towards the edges of the cone. The distribution and kinematics of CO and HCN line emission reveal AGN feedback effects out to $r \sim 500$\,pc, covering the whole extent of the radio jet, with complex kinematics in the molecular gas which displays strong local variations. We propose that this is the result of the almost coplanar jet pushing on molecular gas in different directions as it expands; 
the effects are more pronounced in HCN than in CO emission, probably as the result of radiative shocks.
Following previous interpretation of the redshifted molecular line in the central $5''$ as caused by a molecular outflow, we estimate the outflow rates to be $\dot{M}_{\mathrm{H}_2} \sim 0.9\,M_\odot/\mathrm{yr}$ and $\dot{M}_{\mathrm{dense}} \sim 0.6\,M_\odot/\mathrm{yr}$, which are comparable to the molecular inflow rates ($\sim$1\,$M_\odot/\mathrm{yr}$); gas inflow and AGN feedback could be mutually regulated processes. The agreement with findings in other nearby radio galaxies suggests that this is not an isolated case, and probably the paradigm of AGN feedback through radio jets, at least for galaxies hosting low-luminosity active nuclei.}


\titlerunning{AGN feedback in M51}
\authorrunning{M.~Querejeta et al.}

\maketitle 
\section{Introduction}
\label{Sec:introduction}

\subsection{AGN feedback}
\label{Sec:agnfeedback}

Feedback from star formation (SF) and active galactic nuclei (AGN) plays a key role in reconciling cosmological simulations of galaxy formation and evolution with observations across different redshifts \citep{2015ARA&A..53...51S}. It is often invoked to explain the co-evolution of black holes and their host galaxies \citep{2013ARA&A..51..511K}, 
the mass-metallicity relation 
\citep{2004ApJ...613..898T,2008ApJ...681.1183K}, the bimodality in the colours of galaxies \citep{2015MNRAS.451.2517S}, the enrichment of the intergalactic medium (Martin et al. 2010), and it can prevent galaxies from over-growing in stellar mass \rev{relative to their dark matter halo} \citep[e.g.][]{2006MNRAS.365...11C}. By expelling  molecular gas from the host galaxy, or by changing its ability to form stars, AGN feedback can also regulate star formation; it can either result in the suppression of star formation \citep{2014ApJ...780..186A}, or in its local enhancement \citep{2013ApJ...772..112S}.

AGN feedback is necessary to alleviate the tension between simulations and observations for the most massive galaxies: for $M_* \gtrsim 10^{10}\,M_\odot$ cosmological simulations start to overpredict the stellar mass content of galaxies relative to the mass in their dark matter haloes \citep{2010ApJ...710..903M}. However, feedback is implemented in numerical models in a relatively \textit{ad hoc} way, adjusting its intensity so that the output stellar masses match observations; it remains to be confirmed to what extent these feedback levels are realistic. There is also some observational evidence for the relevance of AGN feedback in quenching star formation in massive $M_* > 10^{10}\,M_\odot$ galaxies \citep[e.g.][]{2016MNRAS.455L..82L}. \rev{The stellar mass of M51, the galaxy that we study here, is $\sim$7$\times 10^{10}\,M_\odot$ \citep{2015ApJS..219....5Q}, and it hosts a low-luminosity AGN. Feedback from the active nucleus might hold the key to the tight connection between bulge and black hole masses \citep[e.g.][]{2009ApJ...698..198G}; in other words, AGN feedback in M51 could be responsible for regulating bulge and black hole growth, preventing the galaxy from developing a massive, young bulge.}

In a cosmological context, a number of studies have recently brought attention to the relevance of AGN-powered outflows \citep[see][for a review]{2012ARA&A..50..455F}; specifically, the last years have seen a plethora 
of detections of ionised ``winds'' driven by AGN activity, including valuable statistics based on large samples \citep[e.g.][]{2014ApJ...795...30B,2016arXiv160104715C}.
There is also ample observational evidence of AGN-driven massive molecular outflows in relatively distant sources 
\citep[e.g.][]{2012A&A...543A..99C,2014A&A...562A..21C,2013A&A...549A..51F,2014A&A...565A..46D,2015A&A...580A..35G};
however, few nearby counterparts have been \rev{resolved}, and the details of feedback are still poorly understood. Notable exceptions AGN-driven include M51 \citep{2004ApJ...616L..55M,2007A&A...468L..49M,2015ApJ...799...26M}, NGC\,4258 \citep{2007A&A...467.1037K}, and since the advent of ALMA, IC\,5063 \citep{2015A&A...580A...1M} and NGC\,1068 \citep{2014A&A...567A.125G}.

Careful analysis of the ALMA $0.5''$ (100\,pc) resolution CO(2-1) data of the radio galaxy IC\,5063 \citep{2015A&A...580A...1M} favours a scenario of a cold molecular gas outflow driven by an expanding radio plasma jet; this results in a high degree of \textit{lateral expansion}, in agreement with numerical simulations of radio jets expanding through a dense clumpy medium \citep[e.g.][]{2011ApJ...728...29W,2012ApJ...757..136W}. However, the high inclination of the disc of IC\,5063 ($i \sim 80^\circ$) complicates the analysis, and does not allow one to directly map the distribution of molecular gas relative to the jet. 
 
A massive (M$_{gas}\sim 3 \times 10^{7}$\,M$_{\odot}$) AGN-driven outflow of dense molecular gas has also been detected by ALMA using high density tracers 
in the inner 400\,pc of NGC\,1068, revealing \rev{complex kinematics} \citep{2014A&A...567A.125G}. These observations suggest that the outflow is efficiently regulating gas accretion in the circumnuclear disc ($r \lesssim 200$\,pc). \rev{Characterising the amount of dense (n(H$_2$)$>10^{4-5}$~cm$^{-3}$) molecular gas that is expelled through this process can help to construct a more complete multiphase picture of AGN feedback.}

From a theoretical perspective, the mere existence of fast molecular outflows is problematic, as large velocities are expected to result in the dissociation of the molecular gas. Cooling into a two-phase medium has been proposed as a mechanism to explain how molecular gas survives outflows (Zubovas \& King 2012, 2014). The numerical simulations from Wagner \& Bicknell (2011) and \citet{2012ApJ...757..136W} also demonstrate the possibility of AGN feedback via radio plasma jets impinging on a clumpy ISM, showing how the interaction between jet and gas \rev{can result in significant lateral expansion (perpendicular to the  direction of propagation of the radio jet)}.

\subsection{The nucleus of M51}

The grand-design spiral galaxy 
M51 constitutes a unique setup due to its proximity  \citep[7.6~Mpc;][]{2002ApJ...577...31C}
and the low inclination of the disc \citep[$i \sim 22^o$;][]{2014ApJ...784....4C}. Its well-studied Seyfert\,2 nucleus \citep{1997ApJS..112..315H,2011AJ....141...41D}
 is seen as two radio lobes that are filled with hot X-ray gas \citep{2001ApJ...560..139T} and an outflow of ionised gas 
\citep{2004ApJ...603..463B}. In the context of the AGN unification picture from \citet{1993ARA&A..31..473A}, the fact that only narrow lines are visible (and no broad lines), which determines the Seyfert\,2 nature of M51, would be explained by obscuration from a dusty torus almost perpendicular to our line of sight; the orientation of the radio jet \citep[inclined 15$^\circ$ with respect to the plane of the disc; ][]{1988ApJ...329...38C}, if perpendicular to the torus, as expected, supports this idea.
We assume that the AGN location is given by the nuclear maser emission position determined by \citet{2015ApJ...815..124H}, RA=13:29:52.708, Dec=+47:11:42.810, which is less than $\sim$0.1$''$ away from the radio continuum peak 
 \citep{1994ApJ...421..122T,2007AJ....133.1176H,2011AJ....141...41D}.

Even if the potential impact of the AGN on the surrounding molecular material has been a matter of debate, high-resolution observations have demonstrated that both CO and HCN are participating in an outflow \citep{1998ApJ...493L..63S,2007A&A...468L..49M}, with an extraordinarily high HCN/CO ratio ($>2$) in the immediate vicinity of the AGN \citep{2015ApJ...799...26M}.

In \citet{2015arXiv151003440Qalt}, we have studied the molecular gas flows across the full disc of M51, probing the transport of gas to the nucleus. Combining our stellar mass map \citep{2015ApJS..219....5Q} with the high-resolution CO gas distribution mapped by PAWS \citep{2013ApJ...779...42S, 2013ApJ...779...43P}, we have found evidence for gas inflow, with rates which are comparable to the amount of outflowing gas ($\sim$1\,$M_\odot$/yr), as we will show.

Our goal is to understand the interplay between nuclear activity and the ISM in the nucleus of M51, relating it to the molecular gas inflow rates that we have already measured. Thus, we present a multi-wavelength study of the inner $\sim$1\,kpc of M51, the region affected by the radio plasma jet.
In order to study the stratification in the response of the molecular gas to AGN feedback in M51, we have obtained new Plateau de Bure interferometric observations of dense gas tracers for the central $60''$ (2\,kpc) of the galaxy. We have detected and imaged three molecules in their $J=1-0$ transition (HCN, HCO$^+$, HNC), but here we will focus on the brightest one, HCN, and compare it to the bulk molecular gas traced by CO from PAWS.
HCO$^+$ and HNC will be analysed in a forthcoming publication.

The paper is structured as follows. We describe the new and archival data used in the analysis in Sect.\,\ref{Sec:data}. The main results are presented in Sect.\,\ref{Sec:results}, and discussed in  Sect.\,\ref{Sec:discussion}. Finally, Sect.\,\ref{Sec:conclusions} consists of a summary, conclusions, and some open questions.

\section{Observations and data reduction} 
\label{Sec:data}

\subsection{Dense molecular gas tracers} 
\label{Sec:densedata}

We have observed the HCN(1–0), HCO$^+$(1–0), and
HNC(1–0) lines in the central $\sim$2\,kpc of M51 with the IRAM Plateau de Bure Interferometer (PdBI) in C and D configurations, and corrected for missing short spacings with single-dish observations from the IRAM 30m telescope (Bigiel et al. submitted). The PdBI observations in D configuration were carried out between 20th August and 13th September 2011 using 5 antennas, with system temperatures between 70 and 130\,K, and precipitable water vapour between 4 and 8\,mm. The PdBI observations in C configuration were performed on 1st and 20th November 2011, and 21st November 2014 with 6 antennas, yielding system temperatures between 60 and 250\,K, and precipitable water vapour between 5 and 8\,mm.
During our observations, the wide-band correlator (WideX) operated in parallel to the narrow-band correlator, simultaneously recording a bandwidth of 3.6\,GHz with a native spectral resolution of 1.95\,MHz (6.6\,km/s), and allowing us to access all of our lines of interest simultaneously. MWC349 was used as the flux calibrator (except for 1st and 20th November 2011, when  1415+463 was used instead). Various quasars were used as the bandpass calibrators for the different sessions (3C84, 3C273, 3C279, 3C345, 3C454.3, 0923+392, 2013+370), and the quasars 1415+463 and J1259+516 were used as amplitude and phase calibrators (except for 20th November 2011, when 1418+546 and 1150+497 were used). The observations were reduced using GILDAS \citep[calibrated with CLIC, and mapped with MAPPING;][]{2005sf2a.conf..721P}.

Single-dish observations of a $4.2'
\times 5.7'$ area ($9 \times 13$\,kpc) were obtained with the IRAM 30m telescope in July and August
2012 (integrating a total of 75 hours). The observations were carried out under typical summer conditions with the 3mm-band EMIR receiver, with a bandwidth of 15.6\,GHz and a spectral resolution of 195\,kHz, and reduced following the standard procedures in CLASS (Bigiel et al. submitted).
We carried out short spacings correction (SSC) of our interferometric data with the single-dish observations from the 30m telescope, which were reprojected to the  centre and channel width of the PdBI observations. The single-dish data were combined with the PdBI observations applying the task \texttt{UV-short} in GILDAS.

CLEANing was performed using the Hogbom algorithm within MAPPING, with natural weighting. This leads to a final \mbox{HCN(1-0)} datacube with a synthesised beam (spatial resolution) of $4.81'' \times 3.94''$ (PA$=71^\circ$), which corresponds to $178 \times 146$\,pc, and a 1$\sigma$ noise level of 0.431\,mJy/beam. We have also performed an alternative cleaning with ROBUST weighting, and we obtain a resolution of $3.66'' \times 3.25''$ (PA$=68^\circ$), with a 1$\sigma$ noise level of 0.460\,mJy/beam. We have chosen a pixel size of $1''$, which produces a cube of $256 \times 256$ pixels and 120 frequency channels of 2.07\,MHz ($\sim$7\,km/s) each.

\subsection{CO data from PAWS} 
\label{Sec:COdata}

We make use of the bulk molecular gas emission probed by the \mbox{CO(1-0)} map of M51 from PAWS \citep{2013ApJ...779...42S, 2013ApJ...779...43P}. The spatial resolution is $1''$ (37\,pc) and covers the central 9\,kpc of the galaxy, reaching a brightness sensitivity of 0.4\,K ($1\sigma$ rms), over 5\,km/s channels.
 Our PAWS map includes short-spacing corrections based on IRAM 30m single-dish data, which by definition recovers all the flux.
For more details on the data reduction, the reader is referred to \citet{2013ApJ...779...43P}.
We note that, for the central region that we are interested in, gas is clearly predominantly molecular 
\citep[HI surface density lower than CO by a factor of more than 10 inside $r < 1$\,kpc;][]{2007A&A...461..143S}.

\subsection{HST archival data} 
\label{Sec:ancillarydata}

We complement the information provided by the molecular gas emission with a number of archival datasets.
We make use of \textit{Hubble} Space Telescope (HST) observations from \citet{2004ApJ...603..463B}, kindly provided to us by the author. This includes a continuum-subtracted [OIII] image from WFPC2/F502N, and a [NII] + H$\alpha$ image from WFPC2/F656N. We note that the WFPC2/F547M image has been subtracted from both images \citep[see][for details]{2004ApJ...603..463B}.

We also utilise HST \textit{Advanced Camera for Surveys} (ACS) mosaics in the F435W (\textit{B}-band), F555W (\textit{V}-band), F658N (H$\alpha$ narrow-band), and F814W (\textit{I}-band), from \citet{2005AAS...206.1307M}. These have been corrected to an absolute astrometric frame of reference as described in \citet{2013ApJ...779...42S}, and we refer the interested reader there for more details.

\subsection{Radio and X-ray archival data} 
\label{Sec:radioXraydata}

The ancillary data that we use also include radio observations with the Very Large Array (VLA), already described in \citet{2011AJ....141...41D}. The whole disc of M51 was mapped at 3.6, 6, and 20\,cm, combining multiple configurations of the interferometer (ABCD for 20\,cm). The map at 20\,cm provides the highest resolution ($1.4'' \times 1.3''$, with a $1\sigma$ rms sensitivity of 11 $\mu$Jy/beam), and we will primarily focus on that image. For comparison, we will also show an X-ray map, for the integrated emission in the 0.3-1\,keV band. The final image is a coadded version of all archival datasets available through the Chandra archive prior to 2011, using \texttt{CIAO} \citep{2006SPIE.6270E..1VF}, and post-processed using adaptive smoothing.

\section{Results} 
\label{Sec:results}

\subsection{Impact of the kpc-scale radio jet on the molecular gas distribution}
\label{Sec:distribution}

\subsubsection{The jet geometry in M51}

\begin{figure*}[t]
\begin{center}
\includegraphics[width=1.0\textwidth]{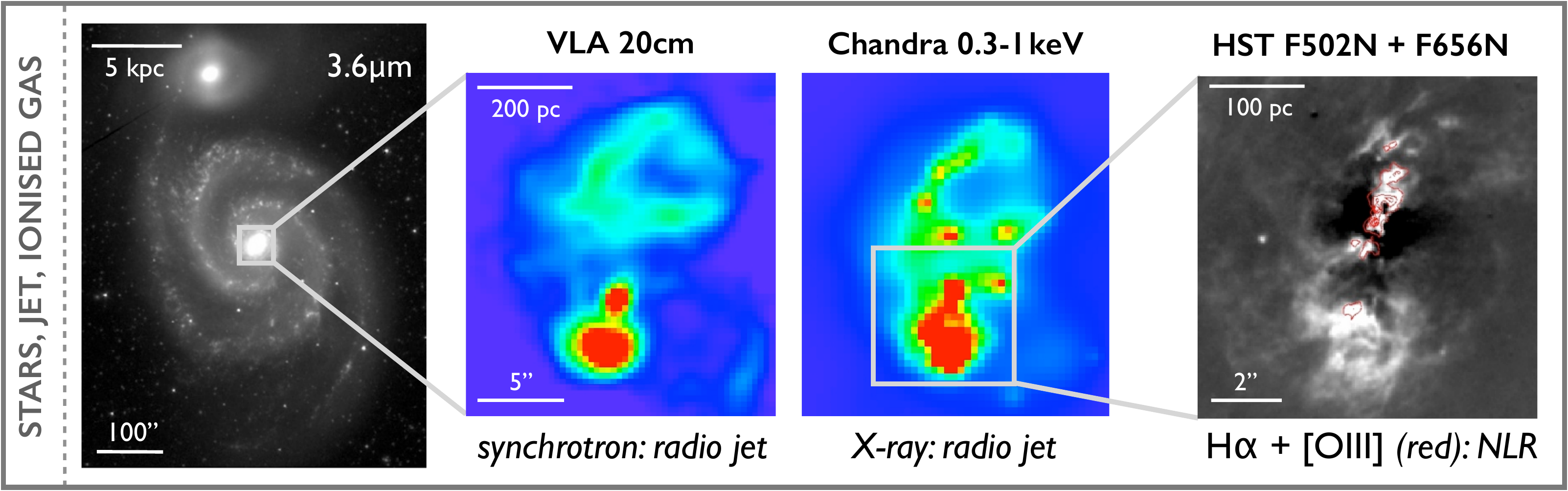}
\end{center}
\caption{\textit{From left to right:} near-infrared image of M51 (\textit{Spitzer} 3.6\,$\mu$m), highlighting the region of interest; VLA 20\,cm continuum image showing the kpc-scale radio jet (Dumas et al.~2011); X-ray emission in the nucleus, from Chandra, integrating emission within 0.3-1 keV; narrow-line region from Bradley et al.~(2004), traced by H$\alpha$ (greyscale image) and [OIII] (red contours).
}
\label{fig:radioXrays}
\end{figure*}

The radio continuum image of M51 (Fig.\,\ref{fig:radioXrays}) shows a number of structures in the inner 1\,kpc, all of them visible at 3.6\,cm, 6\,cm, and 20\,cm \citep{2011AJ....141...41D}. The radio continuum peaks very close to the inferred position of the AGN \citep[RA=13:29:52.708, Dec=+47:11:42.810;][]{2015ApJ...815..124H}. High-resolution VLA imaging by Crane \& van der Hulst (1992) detected the presence of a nuclear radio jet, which extends $2.3''$ (85\,pc) from the nucleus towards the south, with a projected width of $\sim$0.3$''$ (11\,pc); the jet connects to a bright extra-nuclear cloud (XNC) in the south, of $\sim$5$''$ in diameter, which also shows strong optical \rev{emission from ionised gas}. This has been confirmed by \citet{2004ApJ...603..463B}, who also detected a counter-jet in the north, which is curved and shorter ($1.5''$, 55\,pc), and thus is probably the relic of previous nuclear activity. Interestingly, this potential counter-jet points in the approximate direction ($\Delta \mathrm{PA} \lesssim 10^\circ$) of a bright radio component $27.8''$ towards the north. This ``N-component'' is also elongated in the direction of the nucleus, and using a statistical approach, \citet{2015MNRAS.452...32R} have shown that this component is most likely associated with past nuclear activity (and not the chance alignment of a background radio galaxy). A large loop of radio emission is also detected $\sim$9$''$ north of the nucleus (a ring or ``C''-shaped structure), with a diameter of $\sim$10$''$; this structure is in the direction of, but not directly connected to, the northern counter-jet. The radio images from \citet{2011AJ....141...41D} also show a faint southern radio loop that connects to the XNC ($\sim$8$''$ in diameter); to our best knowledge, this southern loop has not been described so far in the literature. On larger scales, the radio continuum emission traces the spiral arms of the galaxy.
In the following, we will loosely use the expression \textit{radio jet} to refer to the radio plasma emission that stems from the nuclear area, which includes the northern loop and the southern XNC, and not only the inner collimated $\sim$4$''$ jet structure. Fig.\,\ref{fig:cartoon}, in which we summarise the main results from this paper, shows an idealised version of the radio plasma structures that we have just described.

\begin{figure*}[t]
\begin{center}
\includegraphics[width=1.0\textwidth]{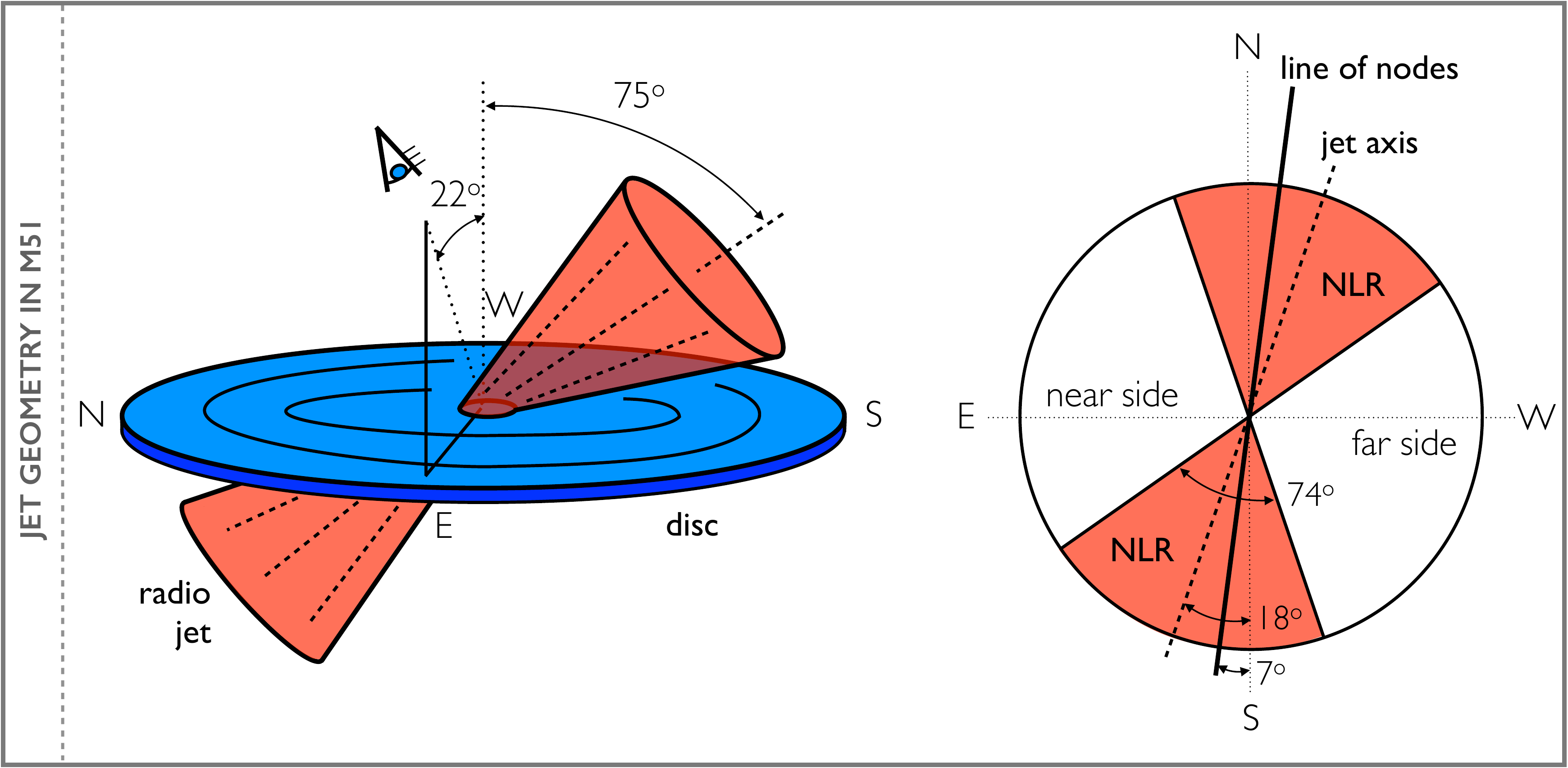}
\end{center}
\caption{\textit{Left:} geometry of the kpc-scale radio jet in M51 relative to the disc and to the observer (the inclination of the jet is inferred from Cecil~1988, and the inclination of the disc, from Colombo et al.~2014). \textit{Right:} orientation and opening angle of the narrow-line region (NLR) from Bradley et al.~(2004), assuming that it is aligned with the jet. We follow Cecil~(1988) and Bradley et al.~(2004) in placing the southern jet closer to us; however, while the inclination of the jet with respect to the disc is well constrained to be low, in the text we discuss the possibility of the northern side of the jet being the one that is closer to us.
}
\label{fig:geometry}
\end{figure*}

From sophisticated modelling of the superposed ionised gas kinematic components in the XNC, \citet{1988ApJ...329...38C} suggested that the (southern) jet forms an angle of $15^\circ$ with respect to the disc of M51 (see Appendix\,\ref{angle}). This inclination of the jet is consistent with the Seyfert\,2 nature of the AGN. Therefore, due to the low inclination of the jet, extended interaction with the ISM is possible. Fig.\,\ref{fig:geometry} illustrates the orientation of the jet and the disc. We note that, even though it has been assumed so far that the southern jet must be closer to us than the northern counterpart (e.g.~Bradley et al.~2004), we think that this is not necessarily true. The fact that the ionised region is more luminous and extended in the south could seem, at first sight, indicative of the southern ionisation cone being closer to us (and therefore, less susceptible to extinction from the optically thick disc). However, the collimated radio jet is longer and brighter in the southern direction (including the southern XNC, which is much brighter at 20\,cm radio continuum than the northern loop); this is probably the consequence of an alternating one-side nuclear activity cycle, perhaps because dense clumps are currently blocking the plasma ejection towards the north of the AGN \citep{2015MNRAS.452...32R}. This could explain why the optical ionised emission lines are stronger in the south, and it does not necessarily imply that the southern jet is closer to us (in fact, if spherical, the XNC would ``block'' the whole molecular thin disc; see Fig.\,\ref{fig:cartoon}).

In Fig.\,2, the radio emission is compared to X-rays. Radio and X-rays trace, to first order, the same structures \citep[see also][]{2001ApJ...560..139T}. The ionised emission mapped through the [OIII] line corresponds to the brightest structures in H$\alpha$, namely those next to the nucleus (inner $\sim$3$''$) and the point of contact of the jet with the XNC, shown to be undergoing an oblique shock \citep{1988ApJ...329...38C}. \citet{2004ApJ...603..463B} showed that the ionised gas along the jet, in the central $\lesssim 3''$, is probably involved in an outflow (although there are important outliers in velocity, which they interpret as gas that is possibly behind the the jet, and participating in a lateral flow away from the jet axis).

\subsubsection{Molecular gas in the vicinity of the radio jet}
\label{Sec:jet}

\begin{figure*}[t]
\begin{center}
\includegraphics[width=1.0\textwidth]{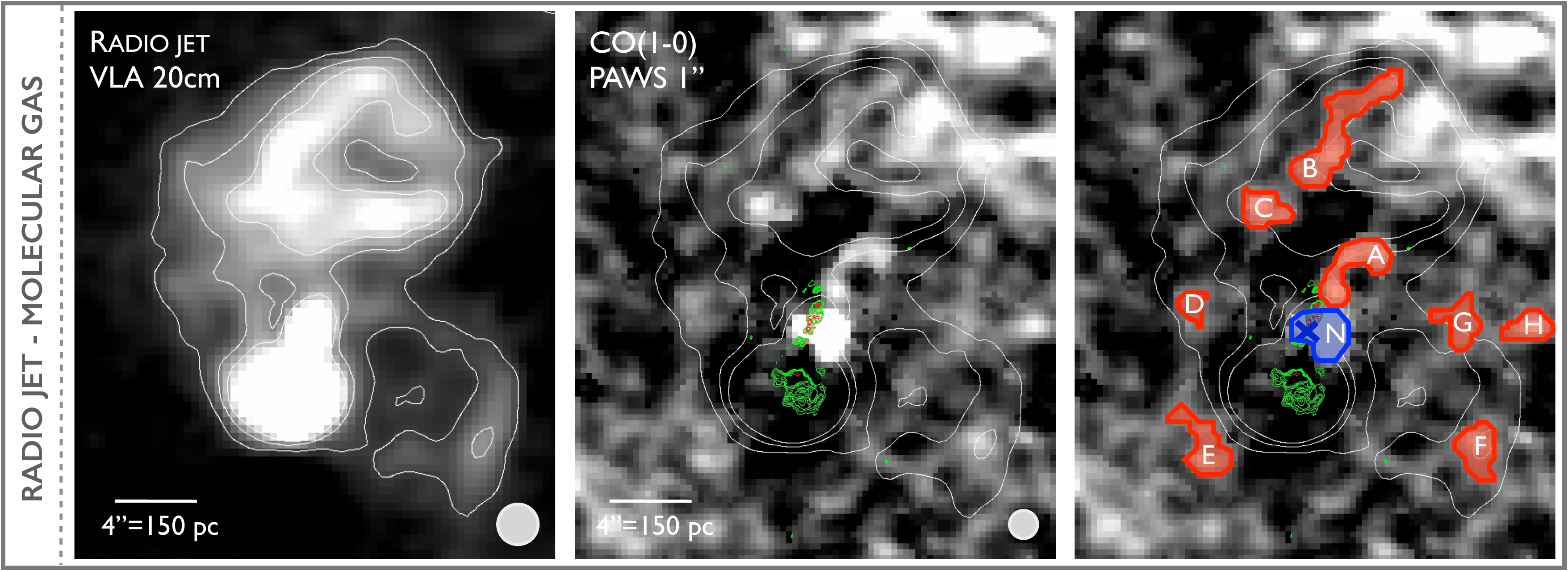}
\end{center}
\caption{\textit{From left to right:} VLA 20\,cm continuum image showing the kpc-scale radio jet in M51; molecular emission in the same region, traced by the \mbox{CO(1-0)} line imaged by PAWS at $1''$ resolution, with green and red contours delineating the NLR (\rev{H$\alpha$ and [OIII] respectively}, see Fig.\,\ref{fig:radioXrays}); same image of molecular emission with some regions identified (quantitatively defined as contiguous regions with CO fluxes above 60\,K\,km/s). \rev{The XNC is the sourthern lobe that becomes apparent in the VLA 20\,cm map, which coincides with strong H$\alpha$ emission (green contours).} The radio image is shown in linear scale from $5 \times 10^{-5}$ to $5 \times 10^{-4}$\,Jy/beam, with three contours at levels 1.5, 3.0, and $4.5 \times 10^{-4}$\,Jy/beam (also reproduced on the maps of molecular emission). The \mbox{CO(1-0)} image is shown in linear scale between 10 and 120\,K\,km/s.
}
\label{fig:radioCO}
\end{figure*}

When we superpose our PAWS \mbox{CO(1-0)} intensity map at our best resolution ($1''$, Fig.\,\ref{fig:radioCO}) on top of the radio emission, an interesting spatial correlation between radio and molecular emission becomes apparent. We highlight as red shaded areas the contiguous regions with \mbox{CO(1-0)} fluxes above 60\,K\,km/s (in the PAWS moment-0 map at $1''$ resolution). 
The curvature of the molecular gas features labelled as A, B, and E agree surprisingly well with the iso-intensity contours from the radio emission tracing the jet. The blobs D, G, and H also seem to accumulate towards the edges of the jet, while C and F overlap with the jet (at least in projection).
 Overall, this gives the impression of CO gas being ``blocked'' by the jet, or alternatively, that the jet is adapting its shape according to the molecular gas distribution, which prevents it from expanding in certain directions. A simple argument of timescales shows that galactic rotation is \rev{probably} much slower than the propagation of the jet, supporting the picture where \rev{the jet expands relative to a quasi-static distribution of molecular clouds}. This agrees with numerical simulations of radio jets propagating through a clumpy molecular medium, in which the jet deforms in an attempt to find the path of minimum resistance \citep{2011ApJ...728...29W,2012ApJ...757..136W}. 
However, this does not mean that the molecular gas is insensitive to the interaction with the jet, as we will show in Sect\,\ref{Sec:kinematics}; \rev{the jet also pushes on the surrounding gas, and} there is probably \textit{reciprocal} interaction between the jet and the molecular gas distribution.

Now, we briefly consider the timescales involved in the AGN feedback process. We can assume that the collimated jet propagates at a velocity close to the speed of light 
\citep[around $\sim$0.9$c$ or even higher, as inferred from observations of other jets and comparison to simulations, e.g.][]{2013ApJ...774...12F}. However, the bubbles and other features of radio plasma that are connected to radio jets expand and dissipate over timescales which are still unclear, or at least largely depend on local conditions \citep{2010MNRAS.406..705B}.

At $R=9''$ (the location of the northern radio loop), the rotation velocity of the molecular gas is 107\,km/s (Meidt et al.~2013), which implies an approximate rotation period of $T_\mathrm{rot} (R=9'')=19$\,Myr; at $2.5''$, where the southern collimated jet hits the XNC, the rotational velocity is 53\,km/s, leading to $T_\mathrm{rot} (R=2.5'')=11$\,Myr. Therefore, even if we assume that the southern radio XNC and the northern loop expand at a moderate speed of $\sim$0.1$c$, their propagation timescales would be of the order of $\sim$10\,000\,yr, three orders of magnitude shorter than the rotation timescales. Therefore, it seems appropriate to consider the picture of a static distribution of molecular gas relative to which the jet and the plasma structures expand; only in the case of very slow plasma propagation ($\ll 0.01 c$) would galactic rotation start to become relevant.

\begin{table}[t!]
\begin{center}
\caption[h!]{Average velocity dispersion in the different phases of molecular gas for the regions labelled on Fig.\,\ref{fig:radioCO}.}
\begin{tabular}{lccccccc}
\hline\hline
  & \multicolumn{7}{c}{FWHM = 2.35$\sigma$ (km/s)} \\   
  & Jet${^a}$ & No jet${^b}$ & Disc${^c}$ &  N & A & B & E \\
  \hline  
 CO(1-0) $1''$  & 28 & 18 & 15 & 68  & 39  & 22  &  21 \\
 CO(1-0) $4''$  &  46 & 24 &  17 & 90 & 57 & 29 &  34 \\
 HCN(1-0) $4''$ & 84  & 44 & 22 & 125 & 82 & 44 &  46  \\
 \hline
\end{tabular}
\label{table:dispersions}
\end{center}
\tablefoot{(\textit{a}) Jet area defined based on 20\,cm VLA image, as the continuous region inside $r < 16''$ with fluxes $F > 0.3$\,mJy/beam.\\
(\textit{b}) Defined as the complementary region to the jet, out to $r < 16''$.\\
(\textit{c}) Region out to $r < 40''$, excluding the jet.}
\end{table}

In Table\,\ref{table:dispersions}, we quantify the velocity dispersions in the different phases of molecular gas for some regions labelled on Fig.\,\ref{fig:radioCO}. We did this by establishing some masks, defined as the positions with CO flux values above 60\,K\,km/s on the $1''$-resolution intensity map (the red shaded regions from Fig.\,\ref{fig:radioCO}), and we calculated the average of the moment-2 map\footnote{The moment-2 maps are the luminosity-weighted standard deviation of the velocity values in each pixel; to avoid being biased by outliers in velocity due to noise, we previously imposed a flux threshold of $5\sigma$ for each velocity channel.} for CO and HCN in those positions. We confirmed that computing luminosity-weighted mean values instead of simple averages leads to very similar results. We also define a \textit{jet} region (fluxes $F > 0.3$\,mJy/beam on the 20\,cm VLA image out to $r < 16''$, before spiral arms start), and the complementary region (\textit{no jet}, $F \leq 0.3$\,mJy/beam on the 20\,cm map). 
For reference, we also calculated a control value as the average velocity dispersion out to $r < 40''$, the limit of the HCN field of view (excluding the jet). 

The velocity dispersions of the molecular gas are clearly higher in the area of the jet than in the same radial range outside the jet (an average of $1.5-2\times$ higher; $2-4\times$ higher when compared to the mean dispersion in the disc out to $r < 40''$). Another important conclusion is that the dense gas traced by HCN shows the largest dispersions, as much as $1.5-2\times$ larger than CO at matched resolution ($4''$); this difference becomes less extreme (only $1.2\times$) when we consider the whole extent of the disc mapped in HCN ($r < 40''$). This is probably indicative of shocks playing a relevant role in the area of the jet and its immediate surroundings. Focusing on the specific regions labelled on Fig.\,\ref{fig:radioCO}, the nucleus (\textit{N}) is the area with the largest line widths (as much as $\sim$100\,km/s at $4''$ resolution). The ridge labelled \textit{A}, which extends north-west from the nucleus, is the second region with largest velocity dispersions; it is located next to the ionisation region, as we will show in Sect.\,\ref{Sec:ionised}. Regions \textit{B} and \textit{E} have similar dispersion values, but still considerably higher than the average disc values (especially in HCN, a factor of $2\times$ higher). The other regions (\textit{C}, \textit{D}, \textit{F}, \textit{G}, \textit{H}), not listed in the table, have dispersions comparable to \textit{B} and \textit{E}, and follow the same trend in terms of HCN being always the broadest line.

\subsubsection{Quantifying the lack of CO in the ionisation cone}  
\label{Sec:ionised}

\begin{figure*}[t]
\begin{center}
\includegraphics[width=1.0\textwidth]{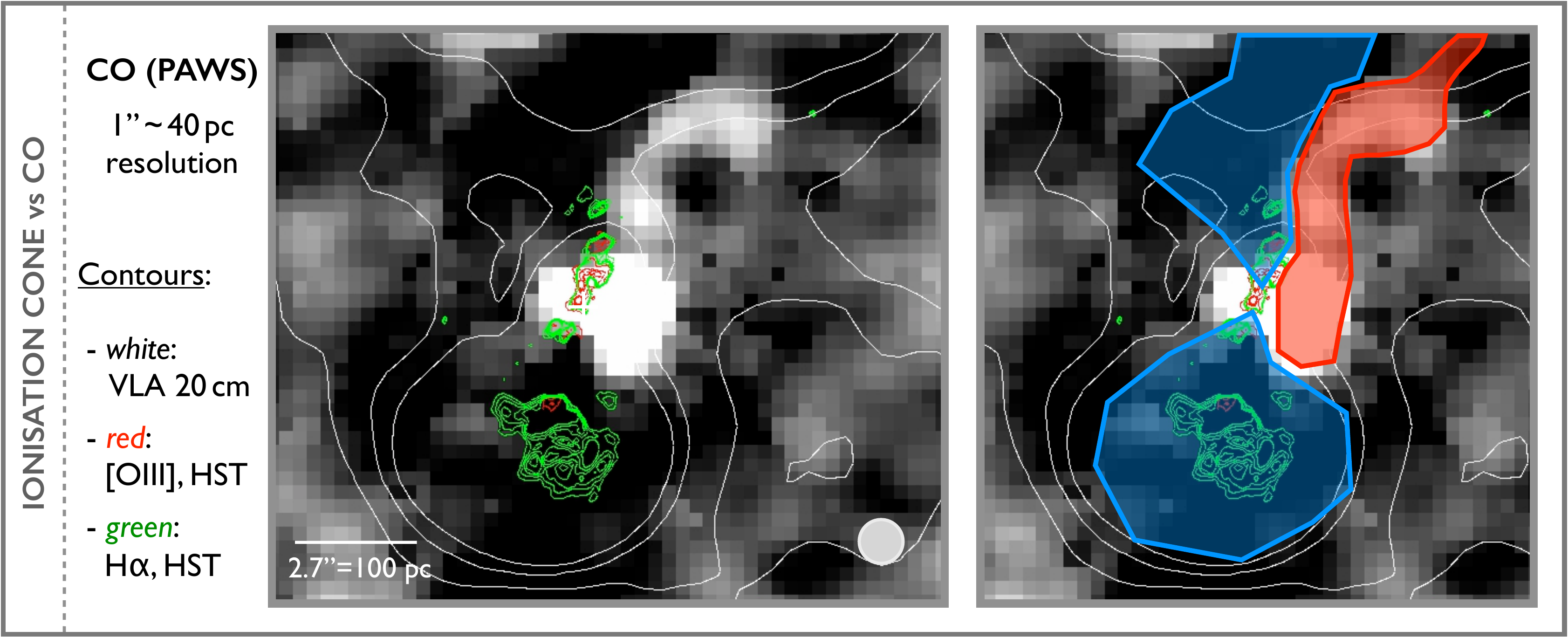}
\end{center}
\caption{Zoom on the central 500\,pc of the PAWS \mbox{CO(1-0)} intensity map (moment-0) at $1''$ resolution, with VLA 20\,cm contours \citep{2011AJ....141...41D}, as well as [OIII] and [NII] + H$\alpha$ from HST, indicating the ionisation region \citep{2004ApJ...603..463B}; the panel on the right highlights how the ionisation region is associated with under-luminous CO emission (blue shaded area), while CO seems to accumulate towards the edge of the jet (red shaded area), supporting and expanding recent findings by \citet{2015A&A...580A...1M}.
}
\label{fig:COhole}
\end{figure*}

In Fig.\,\ref{fig:COhole}, we show a number of contours overlaid on the \mbox{CO(1-0)} intensity map from PAWS (moment-0) at $1''$ resolution. Red and green contours delimit the narrow-line region (NLR), which has an approximate projected biconical shape: in green, H$\alpha$ + [NII],
and, in red, the [OIII] line \citep{2004ApJ...603..463B}.
In white, the radio jet is shown \citep[20\,cm from VLA; ][]{2011AJ....141...41D}, surrounding the XNC that forms part of the ionisation cone ($\sim$5$''$ south from the nucleus), and with the characteristic loop in the north ($\sim$9$''$). This ionisation cone is aligned with the axis of the radio jet, as expected from the AGN unification picture \citep{1993ARA&A..31..473A}.

\begin{figure*}[t]
\begin{center}
\includegraphics[width=1.0\textwidth]{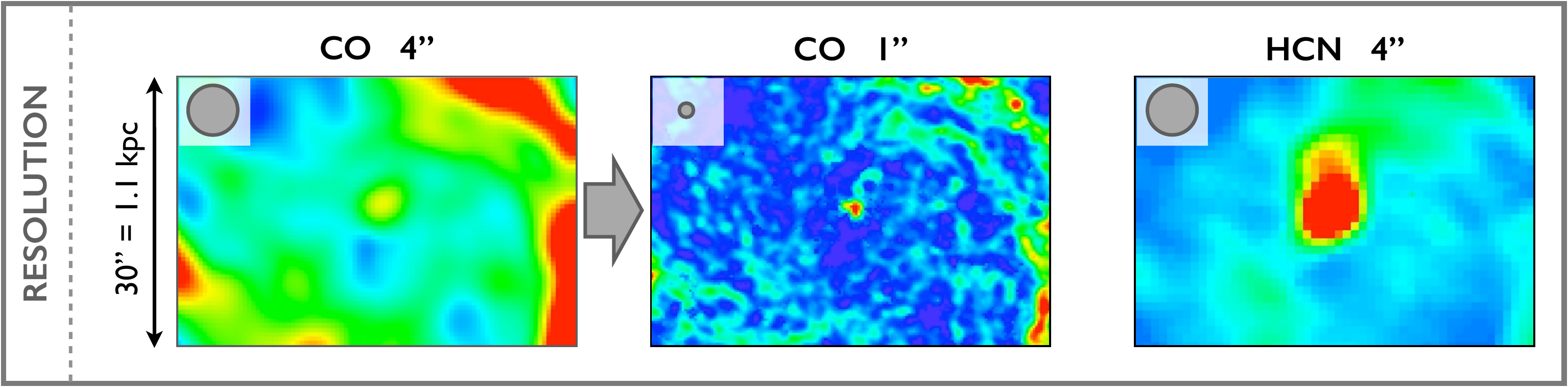}
\end{center}
\caption{Comparison of our PAWS CO intensity maps at $4''$ and $1''$ resolution, and our current best map of HCN at $4''$. This stresses how much the analysis could be hampered by the lower resolution.
}
\label{fig:Resol}
\end{figure*}

We find a relative depression in CO emission coinciding with the ionisation region (Fig.\,\ref{fig:COhole}). While the outflowing ionised gas resembles approximately a bicone, this is not the case for CO (as often assumed for distant unresolved sources); at least in M51, CO accumulates \textit{along the edges of the jet}. Furthermore, we can quantify this scarcity: in the area shaded in blue the mean CO flux is 11\,K\,km/s, whereas the over-concentration of CO emission towards the edge of the ionised region (shaded in red) is 114\,K\,km/s, \textit{a factor of $10 \times$ higher}. Even when comparing to the average flux within the starburst ring (44\,K\,km/s) the blue area is under-luminous by a factor of 4. This
supports the new scenario proposed by \citet{2015A&A...580A...1M} for jet-ISM interaction, in which the jet impinging on a clumpy molecular medium pushes the molecular gas, not only radially, but also laterally.
This important result gets diluted when we consider the PAWS cube at the current resolution of our dense gas tracers ($4''$; Fig.\,\ref{fig:Resol}). With that lower spatial resolution, the difference in CO emission between the jet area and the surroundings is only 41 to 68\,K\,km/s. In fact, a similar mean flux ratio is measured in HCN at $4''$ resolution (12 vs 22\,K\,km/s), so an equivalent scarcity of dense gas might exist in the ionisation region. 

There are a number of reasons that could explain the relative dearth of CO emission in the ionisation cone, including photodissociation by the intense radiation field from the AGN, mechanical evacuation through the expanding plasma, or different excitation conditions. We will discuss these in more detail in Sect.\,\ref{Sec:discussion}.

\subsubsection{Dense and bulk molecular gas distribution} 
\label{Sec:lineratios}

\begin{figure}[t]
\begin{center}
\includegraphics[width=0.48\textwidth]{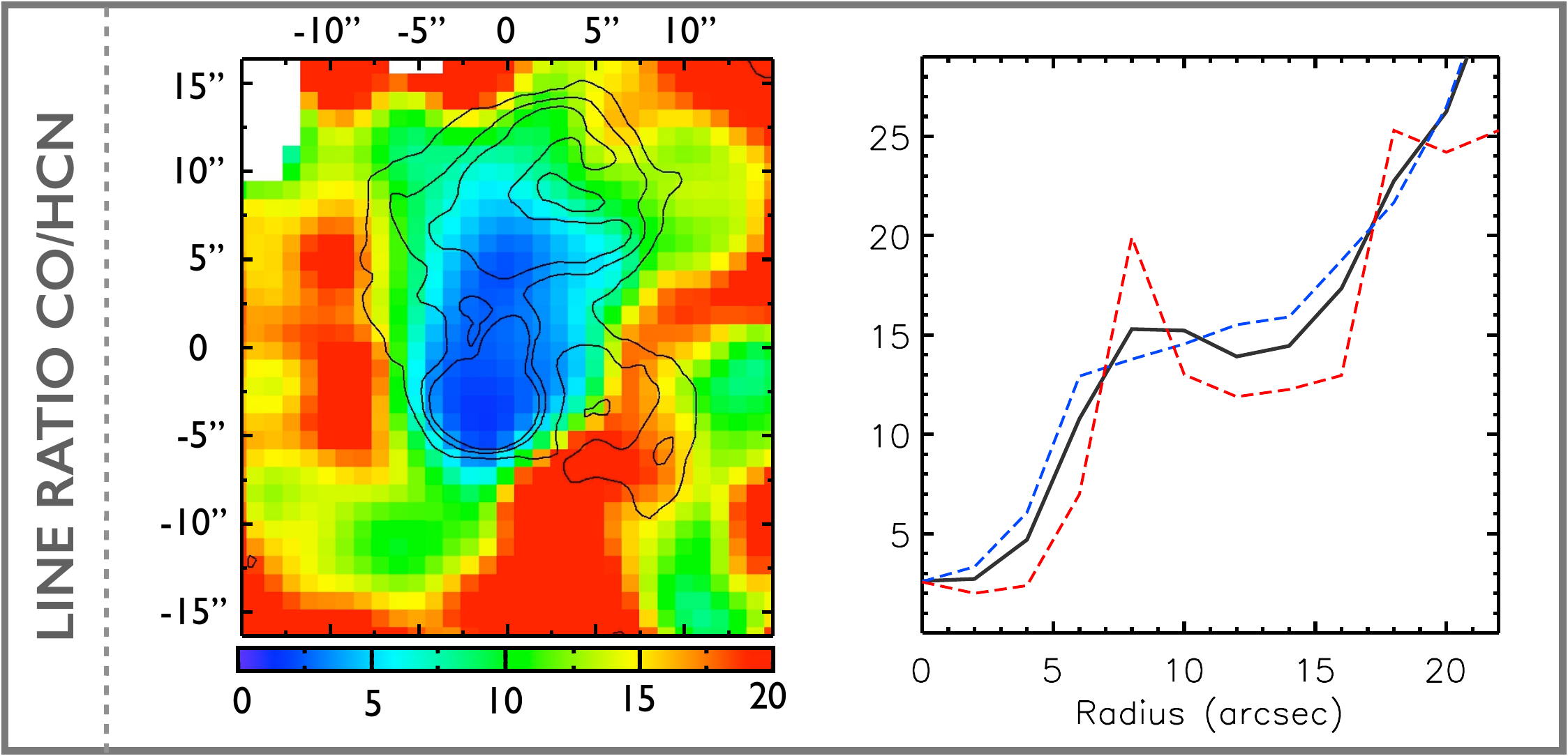}
\end{center}
\caption{Map and radial profiles of the ratio of CO/HCN fluxes. The profile shows the median values of the CO/HCN in radial bins (black line), while the blue dashed line is the ratio of fluxes in the first Gaussian (disc), and the red dashed line is the ratio of fluxes in the second Gaussian (peculiar component); see Sect.\,\ref{Sec:multiplecomp}. 
}
\label{fig:ratioCO-HCN}
\end{figure}

To investigate the dense versus bulk molecular gas distribution in the jet region, we construct a map with the line ratios of the corresponding tracers: \mbox{HCN(1-0)} and \mbox{CO(1-0)}.
Fig.\,\ref{fig:ratioCO-HCN} shows the map of CO/HCN ratio and its radial profile. The line ratio map is calculated by dividing the intensity maps (moment-0) of CO and HCN smoothed to the same spatial resolution ($4''$), as shown in Fig.\,\ref{fig:Resol}, and expressing them in units of brightness temperature~(K).
In agreement with \citet{2015ApJ...799...26M}, we find that the CO/HCN ratio decreases significantly towards the nucleus, reaching \rev{local} values as low as $\sim$1. Matsushita et al.\ suggest shocks and infrared-pumping as the most probable causes of the abnormally high HCN brightness near the AGN position. We probe a much larger area than \citet{2015ApJ...799...26M}, and we find that the line ratio increases almost monotonically as we move towards the edge of our field of view, as expected from the gravitational potential (higher stellar surface density in the centre leads to more dense gas); however, it is interesting to note that there is a plateau and a change in the slope of CO/HCN as a function of radius around $r \sim 8''$ (300\,pc), which could be reflecting a change in the excitation conditions.
This suggests that the radiative effects proposed by \citet{2015ApJ...799...26M} are not restricted to the immediate area connected with the molecular outflow ($r \lesssim 3''$), but might be relevant in the entire region affected by the radio jet. The geometry of the line ratio map also reveals a connection between the \rev{CO/HCN} spatial distribution and the radio jet, although beam smearing effects do not allow us to analyse this in detail.

The dashed lines in Fig.\,\ref{fig:ratioCO-HCN} correspond to the separation into disc and peculiar kinematic components that we describe in Sect.\,\ref{Sec:multiplecomp}. Indeed, as expected, the lowest CO/HCN line ratios are associated with the kinematically peculiar component (red dashed line), while the regularly-rotating disc component typically shows higher line ratios for a given radius.
More molecular transitions are needed to further constrain the mechanisms involved, ideally including isotopologues of CO and other dense gas tracers (for instance, if photodissociation is significant, $^{13}$CO should be even more under-luminous).

\subsection{Kinematics of the molecular gas affected by the radio jet}
\label{Sec:kinematics}

In Sect.\,\ref{Sec:distribution} we demonstrated that there is a spatial anticorrelation between CO emission and the ionisation cone, and also that HCN emission seems to be enhanced in the area covered by the radio jet. This constitutes tantalising evidence of the impact of the radio plasma jet on molecular gas. Here we look for kinematic signatures of this impact in both the CO and HCN line profiles.

\begin{figure*}[t]
\begin{center}
\includegraphics[width=1.0\textwidth]{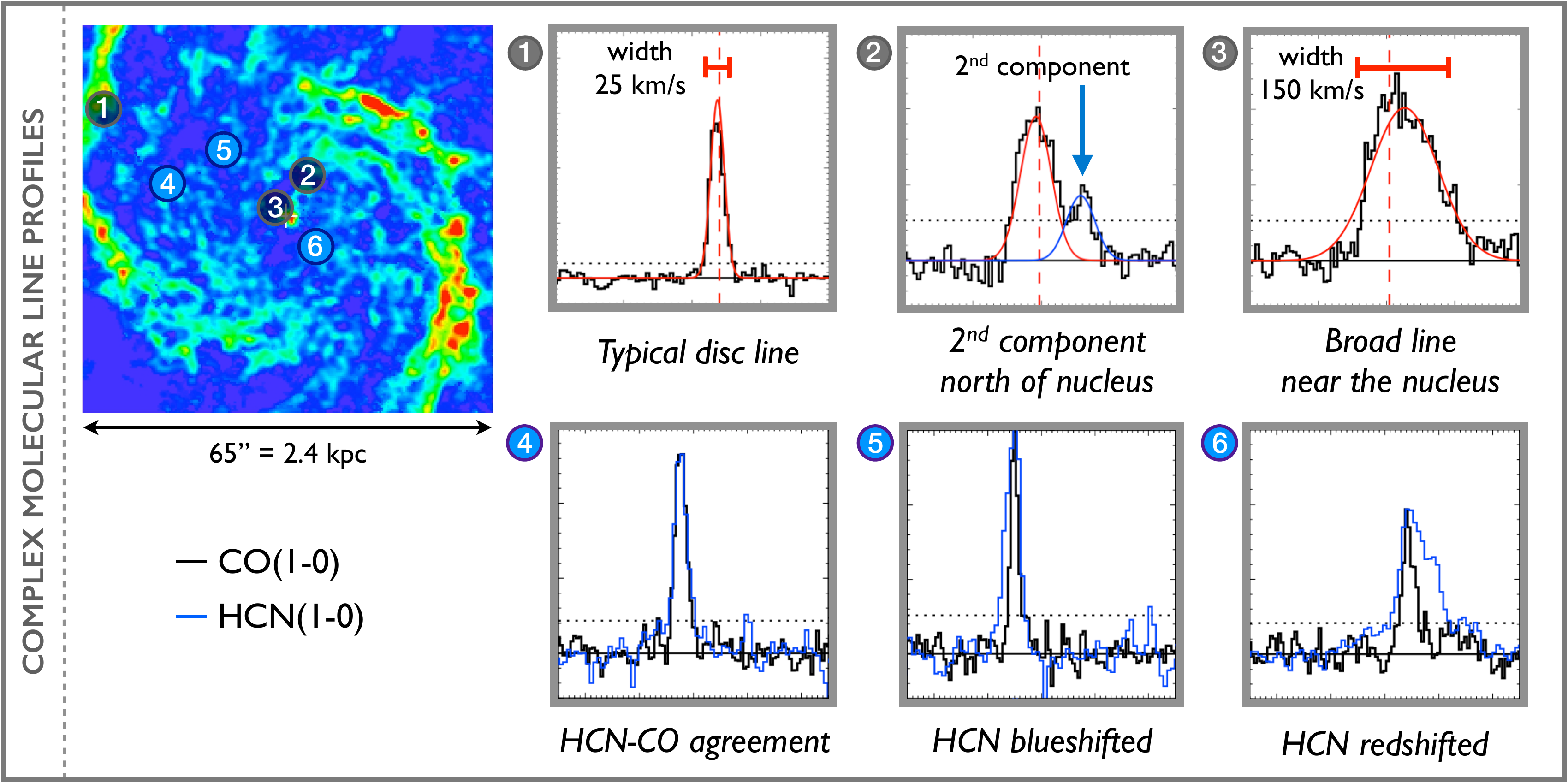}
\end{center}
\caption{PAWS moment-0 map at $1''$ resolution with an indication of the positions where different spectra have been taken. The top row highlights some peculiar components in CO: contrary to the typical CO line of 20-30\,km/s (\textcircled{\tiny{1}}), important secondary components appear as we approach the nucleus (\textcircled{\tiny{2}}) and a very broad line ($\gtrsim 100$\,km/s, \textcircled{\tiny{3}}) in the immediate surroundings of the AGN. The agreement of the main line with the expectation from a \texttt{TiRiFiC} model is remarkable (vertical red dashed line, implementing a Spekkens \& Sellwood bar); 
the distinction between disc and outflow contribution is robust. The bottom row shows some differences between the CO and HCN profiles for three positions: while the agreement is excellent far from the radio jet (\textcircled{\tiny{4}}), there are significant departures as we approach the outer envelope of the jet (\textcircled{\tiny{5}}), and the differences can become extreme as we approach the nucleus (\textcircled{\tiny{6}}). We note that in these line profiles the velocity axes span the range (-250,\,250)\,km/s, with major tickmarks separated by 100\,km/s and minor tickmarks every 10\,km/s; the fluxes are in arbitrary units.
}
\label{fig:lines}
\end{figure*}

Differences between the profiles of CO and HCN for a given position can be used as a diagnostic tool to identify the areas that are affected by mechanical or radiative feedback from the AGN. Some examples are given in Fig.\,\ref{fig:lines}. As we show next in Sect.\,\ref{Sec:HCN-CO}, these relative differences in the CO versus HCN profile shapes become important in the area covered by the radio plasma jet, with a superposition of complex velocity components.

In addition, contrary to the simple profile found across the disc of M51 (FWHM~$ \sim 25$\,km/s), secondary CO and HCN peaks are detected up to $\sim$500\,pc away from the nucleus, and a very broad line (FWHM~$ \sim 150$\,km/s) in the area encompassing the actual AGN (Fig.\,\ref{fig:lines}). These multiple components, identified and discussed in Sect.\,\ref{Sec:multiplecomp}, also point to a \rev{strong} kinematic interplay with the AGN.

\subsubsection{Differences in HCN-CO line profiles} 
\label{Sec:HCN-CO}

In addition to significant local variations in terms of flux (Fig.\,\ref{fig:ratioCO-HCN}), the CO and HCN emission lines also show very important local variations with respect to each other in their line shapes. Even if we scale the CO cube to match the peak brightness temperature of HCN in each pixel, the profiles often diverge. The HCN line is always broader and often skewed with respect to CO in those divergent positions (see Fig.\,\ref{fig:lines} for some examples). Given that HCN traces a denser phase of the molecular gas, and it is more easily excited by shocks and other radiative transfer effects, it is tempting to hypothesise that these divergencies between the CO and HCN lines are revealing a non-linear response of the different molecular gas phases to the feedback from the AGN.

To test this hypothesis, we quantified the difference in the profile shape between the CO and HCN lines as a function of position. For that, we first rescaled the CO intensity map to match the peak brightness temperature of HCN on a pixel-by-pixel basis (at matched resolution, $4''$); then, we obtained a cube with the difference of both profiles, and calculated the moment maps for that HCN-CO cube. To ensure that the moment maps are not driven by noise, we first implemented a threshold cutoff to the HCN and CO cubes, establishing a limit of 5$\sigma$ on each channel in each pixel to identify significant emission. Scaling CO to the peak of HCN or the other way around should lead to exactly the same velocity offsets; however, the relative ``offset intensities'' could in principle change. We have confirmed that scaling HCN to the peak brightness temperature of CO in each pixel, and even rescaling both cubes to an arbitrary peak temperature of 1\,K everywhere, leads to exactly the same velocity (as expected by construction), and very similar ``offset flux'' distribution, with the maximum flux clearly occurring across the jet and leading to an analogous qualitative interpretation. We have also checked against an alternative HCN cube cleaned using robust weighting; in spite of the slightly better resolution ($3.5''$, at the expense of non-uniform noise), the offset maps are virtually identical. We note than the line broadening of HCN(1-0) due to hyperfine structure cannot account for the differences observed with respect to CO(1-0).

\begin{figure*}[t]
\begin{center}
\includegraphics[width=1.0\textwidth]{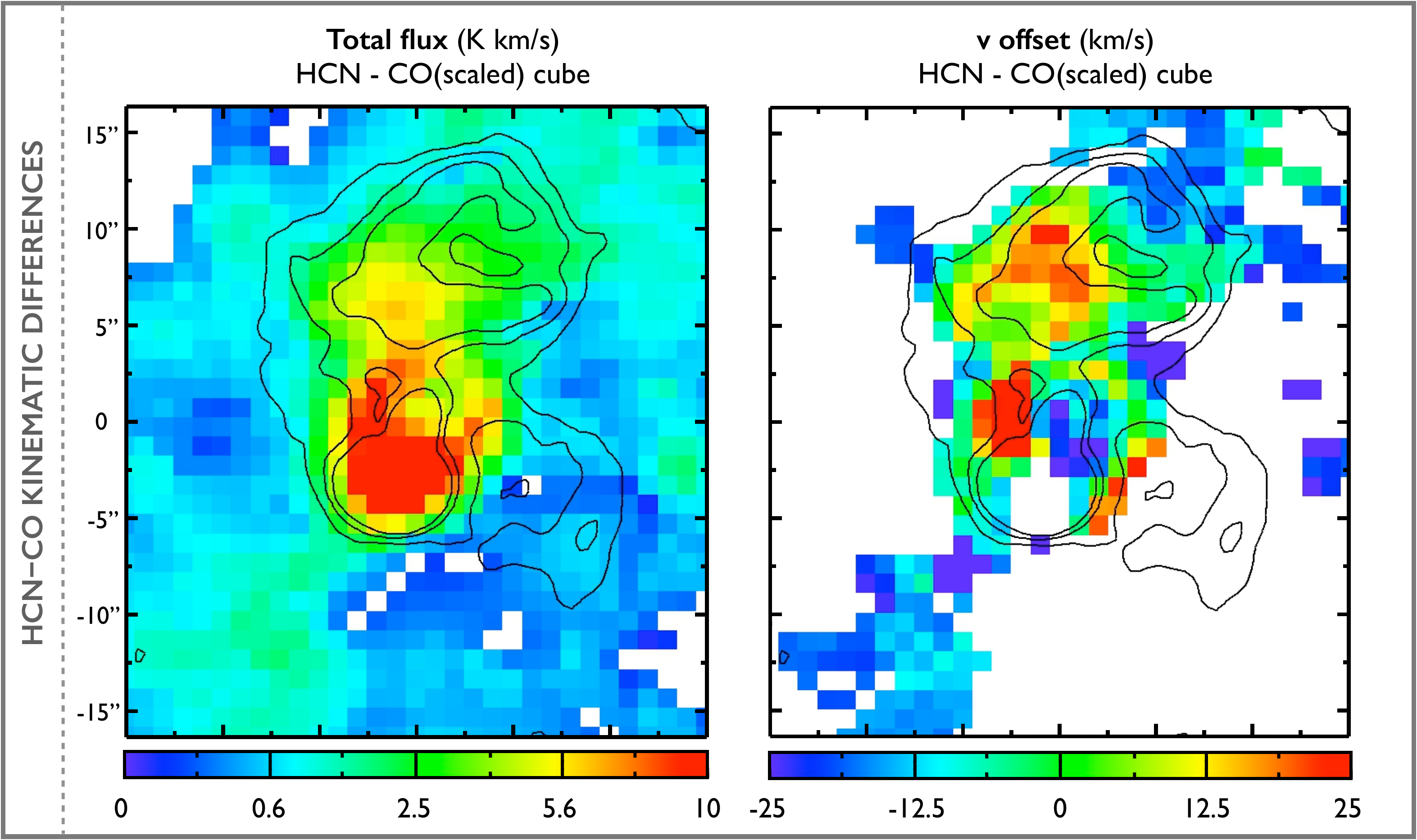}
\end{center}
\caption{Maps showing the relative differences between HCN and CO profiles (when they are scaled to match the same peak brightness temperature). \textit{Left:} total intensity in the HCN-CO scaled cube (moment-0). \textit{Right:} intensity-weighted velocity offset in the HCN-CO scaled cube (moment-1), expressed relative to the CO velocity centroid in each position. The contours show the radio jet from Fig\,\ref{fig:radioCO}, for reference. 
Both maps show the same field of view, $30'' \times 33''$, with large tickmarks indicating a separation of $5''$.
}
\label{fig:HCN-CO}
\end{figure*}

The resulting intensity map (Fig.\,\ref{fig:HCN-CO} left, moment-0) captures the extra emission in the HCN line with respect to CO; it will be zero wherever the HCN profile is a scaled replica of the CO line (same velocity centroid, width, and shape), and non-zero in those positions where HCN diverges from CO, typically because the HCN line is broader and/or skewed. While, in principle, it could happen that negative flux is obtained for some positions (as the CO line could be broader than HCN), we note that this is not significant in the inner $25'' \sim 1\,$kpc.

The velocity map shown in Fig.\,\ref{fig:HCN-CO} (right, moment-1) represents the intensity-weighted average velocity in the scaled CO cube subtracted from HCN. Therefore, this will only diverge from the original CO velocity centroid if the extra emission in HCN is kinematically asymmetric, or in other words, \textit{if the HCN line profile is skewed with respect to that of CO}. This can be expected if the dense gas emission traced by HCN is preferentially associated with energetic and dissipative phenomena such as shocks. However, due to our limited resolution, it can also happen that multiple kinematic components blend together for a given line of sight (by the $\sim$4$''$ beam), and this could result in zero bulk velocity offset with respect to CO due to mere superposition; however, even in this situation, the HCN profile will show wings with respect to CO on both sides (the HCN line will have overall larger line width), which will result in detectable extra emission in the HCN-CO intensity map (Fig.\,\ref{fig:HCN-CO} left). For clarity, in the final velocity map we blank all pixels where this integrated extra emission is less than $1\,$K\,km/s (referred to the HCN flux scale).

At first glance, what is most remarkable about the
HCN-CO line profile differences in Fig.\,\ref{fig:HCN-CO} is that the highest extra flux in the central area closely follows the structure of the radio plasma jet. While there are holes near the very centre (because CO is extremely under-luminous in the ionisation cone, see Sect.\,\ref{Sec:ionised}), the areas of large HCN-CO offsets coincide with strong radio continuum emission, particularly towards the edges of the northern loop, the sides of the radio continuum peak (AGN position), and the edges of the southern XNC. Extended HCN emission (with significant differences with respect to CO, and with a high velocity dispersion) are also found towards the north-west and south-east of jet, aligned with the radio plasma structures. Overall, this is indicative of HCN emission being affected by the radio jet out to distances of at least $\sim 12'' \sim 450$\,pc.

In spite of this strong spatial agreement, kinematically the situation is more complex. 
The right panel of Fig.\,\ref{fig:HCN-CO} shows very strong local variations, typically covering the range \mbox{(-30,\,30)\,km/s}.
 Even though these velocities are not extremely high, we note that they could be reflecting much larger underlying velocity components, because: (a) the moment-1 map shows the intensity-weighted average of the velocities, to the point that negative and positive velocity offsets can cancel each other; and (b) since M51 is seen almost face-on, any lateral planar motions will project to the line of sight as small velocity components. For example, typical moment-1 values around 20\,km/s in the northern loop correspond to HCN profiles which have significant emission kinematically offset by more than 50\,km/s, which accounting for projection effects could translate into intrinsic velocities above 100\,km/s.
 
\begin{figure*}[t]
\begin{center}
\includegraphics[width=1.0\textwidth]{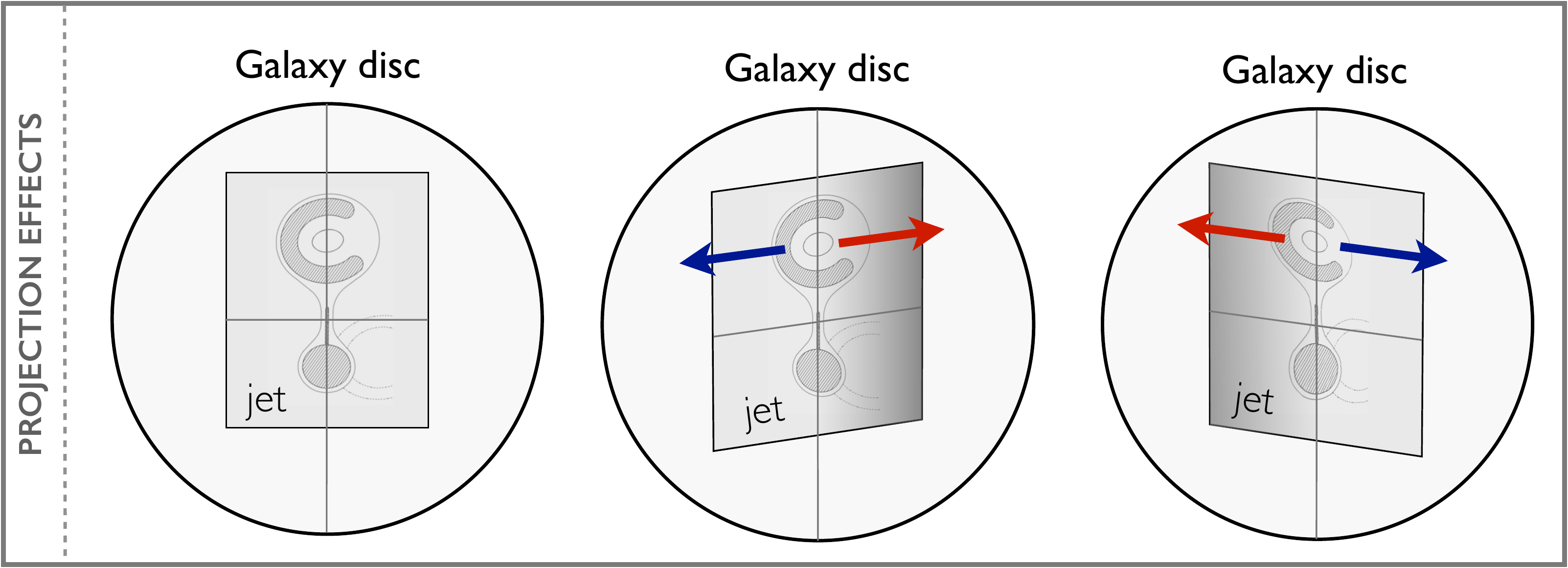}
\end{center}
\caption{Cartoon illustrating how a relatively small tilt between the plane of the jet-loop structure and the disc of M51 can result in net redshifted or blueshifted velocities on alternating sides of the galaxy (because M51 is almost face-on). The differences between HCN and CO kinematics in Fig.\,\ref{fig:HCN-CO}, with hints of redshifted material in the east of the northern loop and blueshifted in the west, could be indicative of the situation suggested by the right panel of this cartoon.
}
\label{fig:orientation}
\end{figure*}

In M51, the line of nodes coincides approximately with the north-south axis (PA$=-7^\circ$), and the radio jet axis is also not far from the line of nodes (PA$=-18^\circ$). 
Even though the inclination of M51 is not very large ($i = 22^\circ$), the near side of the galaxy is the \textit{east side} (left in our figures; \citealt{2014ApJ...784....4C}).
Any motions parallel to the line of nodes will have no projected components along our line of sight; therefore, we are essentially insensitive to radial propagation along the jet from the perspective of kinematics. On the other hand, any motions of lateral expansion associated with the jet will result in detectable projected velocity components.
Assuming that the jet propagation is coplanar with the disc of M51, we would expect preferentially blueshifted components in the east (the near side, pointing towards us), and redshifted in the west (the far side, directed away from us). However, this is not what we see in the right panel of Fig.\,\ref{fig:HCN-CO}, at least in the northern loop; we find preferentially redshifted velocities in the east, and blueshifted in the west. This apparent contradiction can easily be explained if the plane of the jet is tilted with respect to the midplane of the molecular disc; this could easily result in the observed kinematics as the consequence of lateral expansion, as illustrated by Fig.\,\ref{fig:orientation}.

Assuming that the northern loop is a relatively flat structure, the complex kinematics could also be partially explained as the result of motions \textit{perpendicular} to the plane of the galaxy: the loop could be pushing the molecular gas \textit{up and down} in the galactic disc, and not only laterally in the plane of the galaxy. In any case, the resulting projected velocities will largely depend on the relative position of the affected molecular clouds or clumps vertically, relative to the galaxy midplane, which can contribute to the observed HCN-CO velocity distribution.

In general, the kinematics of HCN-CO could be reflecting at least two different possibilities: (a) the response to the radio jet traced by the difference between HCN and CO line shapes is highly stochastic, with intrinsically large variability with position;
(b) we see different red- and blueshifted components superposed in the same line of sight due to our limited resolution ($\sim$4$''$), because the (possibly coherent) structures involved are close to each other in projection, and are blended by the beam.
Even though our limited resolution does not allow us to distinguish between (a) and (b), in both cases the data would imply significant lateral expansion.
The data conclusively proves that the molecular gas is affected by the AGN through the radio jet to distances of at least $\sim$12$''$ (450\,pc) in the north, and $\sim$7$''$ (260\,pc) in the south, with rapid phase changes or different excitation conditions, which involve deprojected velocity offsets between HCN and CO of $\gtrsim 100$\,km/s.

Even though we cannot resolve HCN into small structures ($< 4''$) with our current dataset, the rapid local variations in the velocity offset with respect to CO make us hypothesise that the HCN emission comes from a clumpy medium (at least as clumpy as the bulk molecular gas traced by CO). This possibility should be confirmed with data at higher resolution.

\subsubsection{Identification of multiple Gaussian components in molecular line profiles} 
\label{Sec:multiplecomp}

In Fig.\,\ref{fig:lines} we show some examples of the differences in CO line shape for three different positions in the central area of M51. In addition to significant offsets between HCN and CO, many positions show evidence for multiple components.
With the goal of characterising those multiple components, we perform a kinematic separation of different Gaussian contributions, on a pixel-by-pixel basis, for each of the cubes (CO, HCN). For each line profile (i.e. each pixel), we iteratively fit 1, 2, or 3 Gaussians, minimising $\chi^2$.
 In the fitting process, we use the peak velocity as the starting fitting point for the first Gaussian, whereas for the second and third, we allow the fits to start from a number of equispaced spectral positions (a total of 20 starting points) that cover the whole velocity axis. 

To determine which Gaussian fits are significant, we rule out \textit{a posteriori} any fitted Gaussians which correspond to a region of the profile which does not have \textit{at least} three adjacent channels above 5$\sigma$ (the rms noise level\footnote{We calculate $\sigma$ as the standard deviation of the signal in line-free channels: channels 5--29 for PAWS-CO (195\,km/s\,$<v<$\,315\,km/s); channels 5--38 for HCN (84\,km/s\,$<v<$\,315\,km/s).}). This is to avoid spurious fits to noise (since noise is spatially correlated, but fully uncorrelated among adjacent channels). 
This is an extreme threshold, and it is chosen to avoid potential degeneracies due to fitting multiple Gaussians when we are dominated by noise (Sect.\,\ref{Sec:discussion}), and to highlight the most significant components. 
Even when the emission itself follows a perfect normal distribution, if noise is not negligible compared to the peak of the Gaussian, it can happen that a random dip in the profile which originates from abrupt channel-to-channel noise variations will result in $\chi^2$ being minimised by fitting two Gaussians instead of one; therefore, the kinematic decompositions performed this way face some practical difficulties, and that is why we prefer to show maps with a very high threshold.
We verified that lowering this threshold does not qualitatively change our conclusions.

About 30\% of the pixels with significant CO emission at $r \lesssim 1$\,kpc are better described by multiple Gaussian components than a single Gaussian profile. The equivalent fraction for the HCN cube is 22\%. 
We make the conservative assumption that the Gaussian that carries most flux is the one associated with the disc. We confirm that this is a reasonable hypothesis with a \texttt{TiRiFiC} synthetic kinematic model that implements a rotating disc and a \citet{2007ApJ...664..204S} bar for M51 (see Appendix\,\ref{sec:TiRiFiC}); only in the very centre ($r \lesssim 5'' = 200\,$pc) do significant divergencies ($\Delta v > 10$\,km/s) appear between the centre of the first Gaussian and the velocity centroid from \texttt{TiRiFiC}. The divergencies in the inner $r \lesssim 5''$ are expected, as this is the region where the line profile becomes extremely broad, and it starts to become impossible to kinematically separate disc from outflow (Fig.\,\ref{fig:lines}). In Appendix\,\ref{sec:disccontr}, we estimate the maximum possible contribution from a hypothetical disc component in this central region, and we will include this upper limit for $r \leq 5''$ in the maps and analysis from now on.

\begin{figure}[t]
\begin{center}
\includegraphics[width=0.48\textwidth]{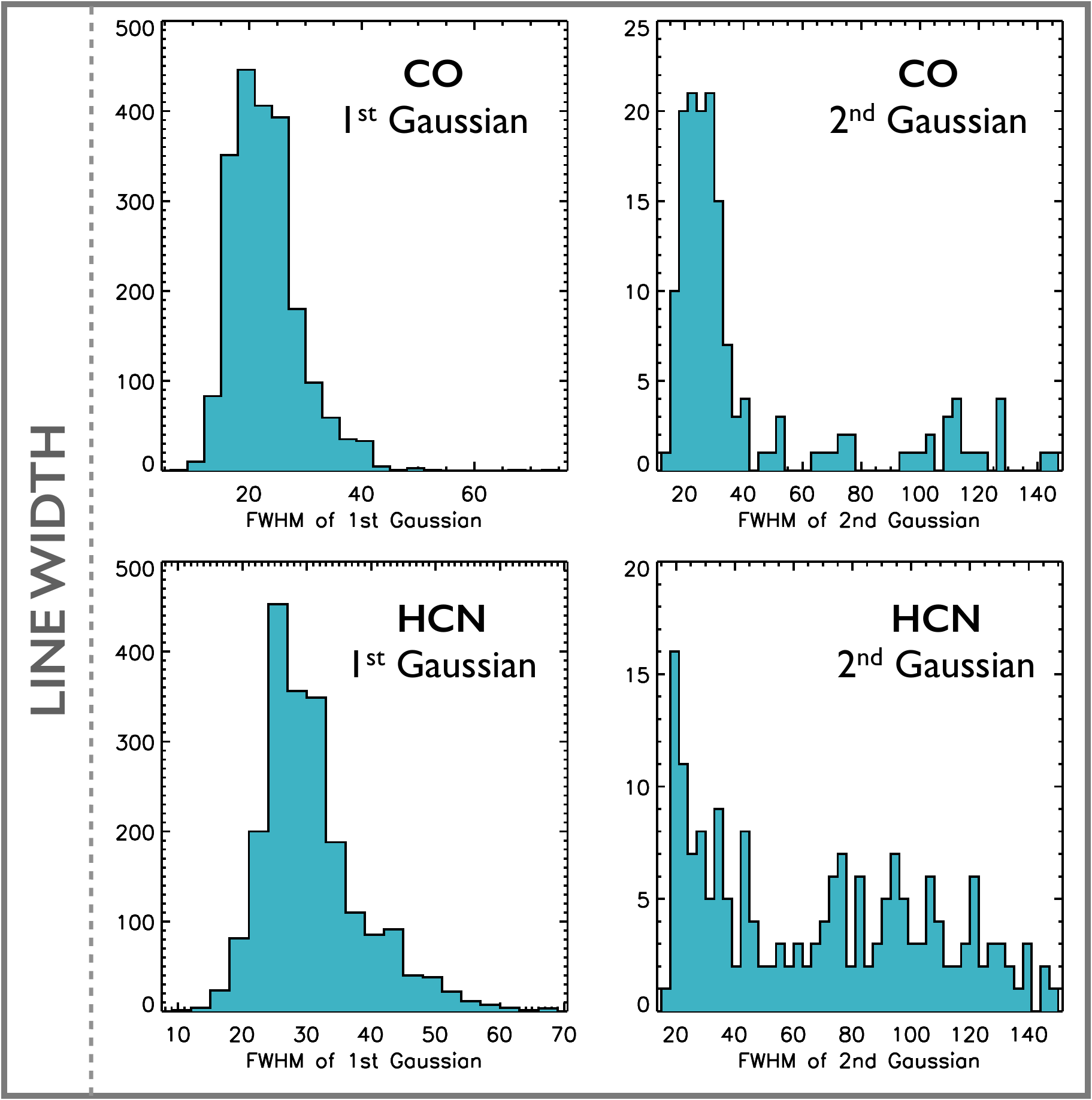}
\end{center}
\caption{Histograms showing the width of the fitted Gaussians, for CO (top) and HCN (bottom). The left histograms show the distribution of widths for the first Gaussian, which follows regular motions as expected for a rotating disc, while the right histograms display the width distributions for the second Gaussian, which show significant departures from the velocity expectations for the disc.}
\label{fig:width}
\end{figure}

\begin{figure*}[t]
\begin{center}
\includegraphics[width=1.0\textwidth]{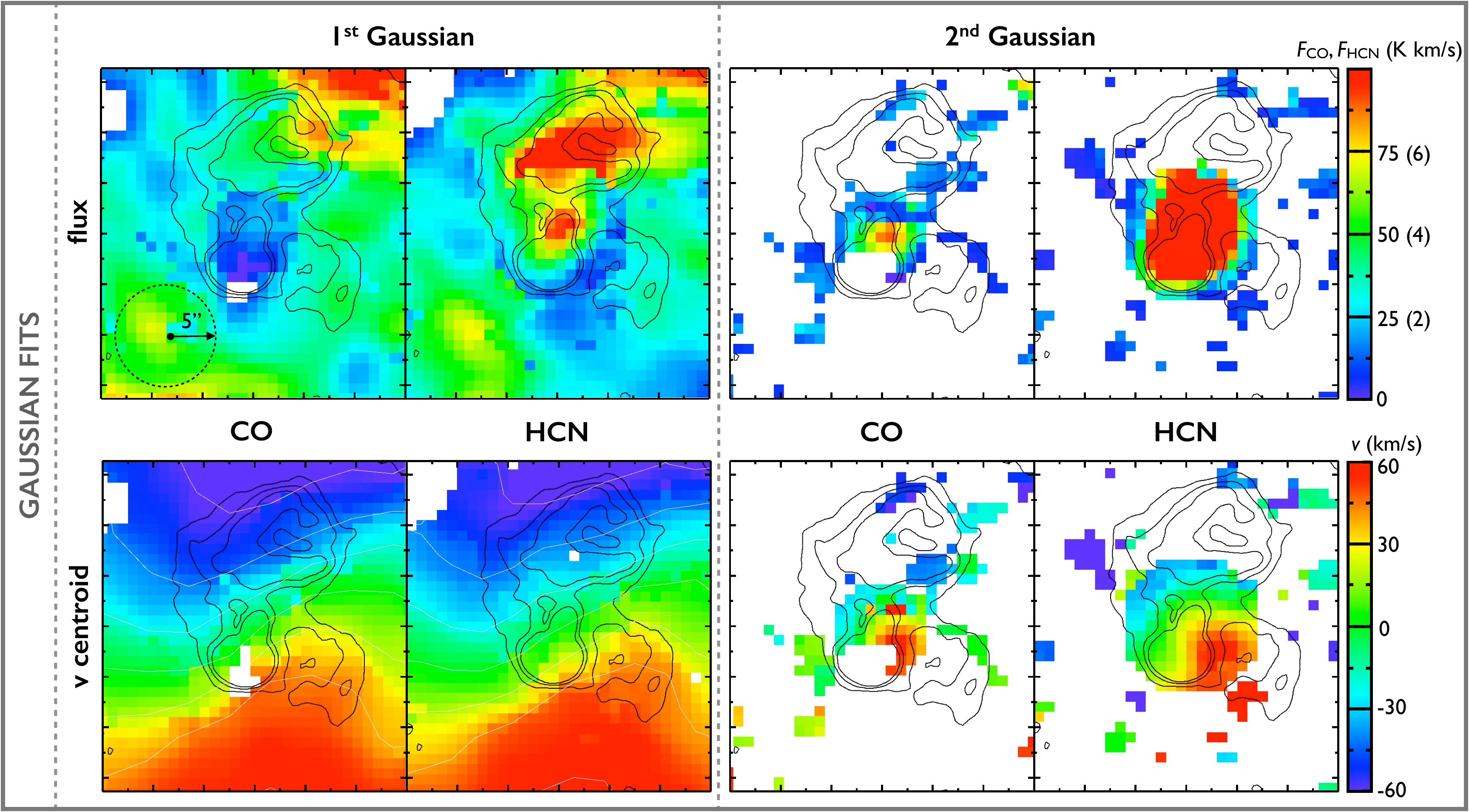}
\end{center}
\caption{Flux and velocity centroid of the Gaussians fitted on a pixel-by-pixel basis to the CO and HCN profiles (Sect.\,\ref{Sec:multiplecomp}). 
The top row shows the integrated flux of the first (main) and second Gaussians, while the bottom row shows the corresponding velocity centroids. The white isovelocity contours displayed on the velocity maps cover the range [-60, 60]\,km/s, in intervals of 20\,km/s. The colourbar and scales covered are shown in the right (in parenthesis for HCN).
All maps show the same field of view, $30'' \times 33''$, with large tickmarks indicating a separation of $5''$. For reference, the dashed circle on the top-left panel has a radius of $5''$ (185\,pc).
}
\label{fig:GaussianMom}
\end{figure*}

\subsubsection{Results of the decomposition into Gaussians}
\label{Sec:resultsmultiplecomp}

As demonstrated by Fig.\,\ref{fig:width}, the main CO line associated with the disc has a very restricted range of widths, typically 20\,km/s\,$\lesssim$\,FWHM\,$\lesssim$\,30\,km/s; this line is in agreement with the idealised model we constructed with \texttt{TiRiFiC}, presented in Appendix\,\ref{sec:TiRiFiC}. When we look at the width of the main line (the one that carries most flux) we observe a very strong bimodality:
in the central $\sim$5$''$, we find an extremely broad line (FWHM\,$\lesssim$\,150\,km/s), whereas outside that region, its width is more moderate (but still larger than the typical disc line). Additionally, for many positions, we identify multiple Gaussian kinematic components, following the technique described in the previous section. The width of these second components typically ranges from $\sim$10\,km/s to $\sim$40\,km/s in CO, with many more high-dispersion outliers than in the first Gaussian; the widths in HCN show an even more uniform distribution, covering the whole range $20-140$\,km/s almost uniformly (see Fig.\,\ref{fig:width}).

For the molecular emission that departs from the disc expectations, there is also a clear dichotomy in terms of flux: out to $5''$, the second component, presumably associated with the outflow, clearly dominates in flux ($> 5\times$ larger than the upper limit for the disc contribution), and it shows up as a coherent, continuous structure (Fig.\,\ref{fig:GaussianMom}). Outside $5''$, however, the peculiar components appear as small islands, probably associated with specific molecular clouds or cloud complexes. This is not too surprising if, instead of a coherent entrainment of the molecular material by the AGN, the effects we detect result from the radio jet pushing the gas in the disc laterally as it expands. In that case, due to projection, we will find different kinematic components (in some parts blueshifted, in others, redshifted) much more patchy in nature. This reflects a similar situation to what we interpreted from the HCN-CO velocity differences (Sect.\,\ref{Sec:HCN-CO}).

Specifically, in the areas where gas seems to accumulate next to the ionised region (for instance, those shaded in red in Fig.\,\ref{fig:COhole}), we find multiple peculiar components, which change very rapidly with position (this becomes visible in the CO cube at $1''$ resolution, where changes are of the order of $1''$, potentially indicative of shocks). For example, the faint radio loop in the south, which connects to the XNC on its western side, seems to be associated with a coherent secondary component in CO, which extends $\sim$5$''$ parallel to the northern limit of the loop.
 This region does not correspond to a significant HCN-CO profile difference, though, and does not appear in the maps of Fig.\,\ref{fig:HCN-CO} (this is because, in fact, HCN also shows this secondary blueshifted component, but it is not above the high S/N threshold imposed on the maps of Fig.\,\ref{fig:GaussianMom}).

Overall, this also allows us to refine our calculation of the amount of gas that shows significant departures from planar disc motions. By adding up the contributions from the second and third Gaussians over each of the pixels, we arrive at a fraction of 95.5\% for CO in the inner $20''$ due to peculiar kinematic components, and as much as 99.9\% in the inner $5''$ (for HCN, these values are 29.7\% and 79.6\%, respectively).  This is mostly driven by the large amount of flux that deviates from the expectations for regular rotation in the inner $\sim$5$''$.

\subsection{Complex structures at the base of the jet} 
\label{Sec:baseofjet}

Now we focus on the central $3''$ (110\,pc) of M51 to probe the effects of feedback in the immediate vicinity of the AGN. We use position-velocity diagrams for CO and HCN (Sect.\,\ref{Sec:pvdiagrams}) to verify that our data are consistent with the molecular gas outflow proposed by \citet{2007A&A...468L..49M}, and to estimate the molecular gas and dense gas outflow rates. We also present evidence for an intriguing structure with very blue optical colours in the central $1''$.

\subsubsection{Dense and bulk molecular gas outflow rates} 
\label{Sec:pvdiagrams}

\begin{figure}[t]
\begin{center}
\includegraphics[width=0.48\textwidth]{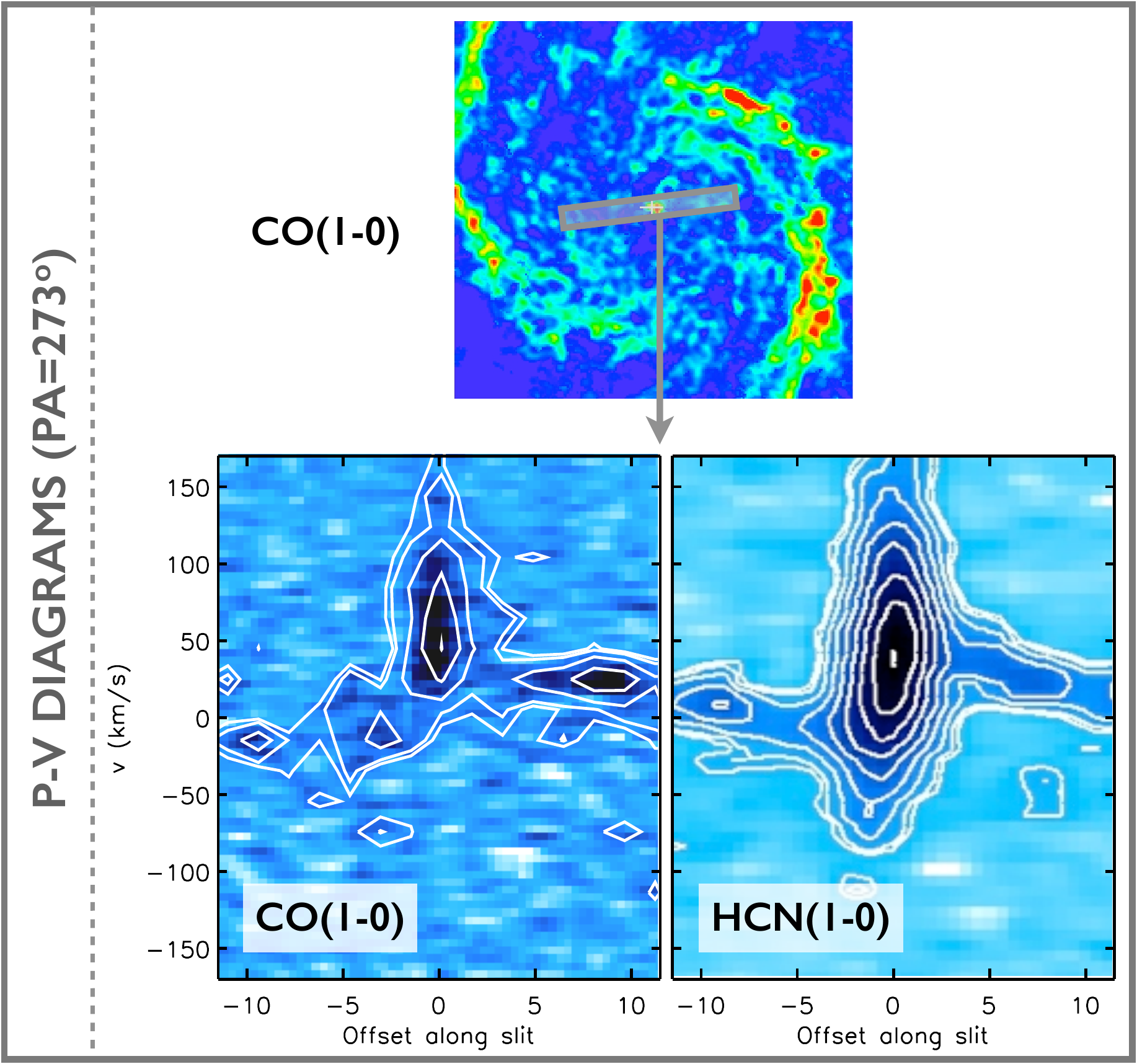}
\end{center}
\caption{Integrated intensity map of our PAWS (PdBI Arcsecond Whirlpool Survey) \mbox{CO(1-0)} cube at $1''$ resolution, with an indication of the slit used for the p-v diagrams shown next. Position-velocity diagrams extracted along a PA$=273^\circ$, for CO ($1''$ resolution) and HCN ($4''$), in which the redshifted wings of the outflow become manifest (with a blueshifted counterpart hinting in HCN). 
}
\label{fig:pv}
\end{figure}

Position-velocity diagrams are one of the most straight-forward tools to probe departures from regular motions in galaxies; when taken along the \rev{local} kinematic minor axis, an approximately flat profile is expected (constant projected velocity), while any orthogonal velocity components will stand out as outliers in the profile (e.g. outflows).
In M51, such departures from regular rotation in the
molecular gas become clear in the position-velocity plots of Fig.\,\ref{fig:pv}, with redshifted components reaching $v \sim 150$\,km/s in both the bulk molecular gas identified through CO emission and the dense gas traced by HCN.

Under the simplistic assumption of a conical outflow (see \rev{caveats below}), we can follow \citet{2010A&A...518L.155F} and \citet{2012MNRAS.425L..66M} to obtain an estimate of the outflow rates. In the case of a uniformly filled (multi-)conical outflow:

\begin{equation}
\frac{\mathrm{d}M}{\mathrm{d}t}= 3 \times V_\mathrm{out} \times M_\mathrm{gas}/ R_\mathrm{out} \times \tan (\alpha),
\end{equation}  \label{eqn:outflow}

\noindent
where $M_\mathrm{gas}$ is the total molecular gas involved in the outflow, $V_\mathrm{out}$ and $R_\mathrm{out}$ are characteristic values of the velocity and radius of the outflow, and  $\alpha$ is the inclination angle of the outflow with respect to the \rev{line of sight}.
This is more conservative (it results in lower outflow rates) than assuming a \rev{spherical (possibly fragmented)} shell-like geometry, which would correspond to:

\begin{equation}
\frac{\mathrm{d}M}{\mathrm{d}t}= V_\mathrm{out} \times M_\mathrm{gas}/ dR_\mathrm{out} \times \tan (\alpha),
\end{equation}
\\
where $dR_\mathrm{out}$ is the thinness of the shell instead of the characteristic radius \citep[and, in general, $dR_\mathrm{out} \ll R_\mathrm{out}$;][]{2012MNRAS.425L..66M}.

\rev{In fact, the spatial distribution of the gas that shows extreme velocities near the centre does not quite support either the biconical nor the spherical shell scenarios; this is also confirmed by the higher-resolution CO maps from \citet{2007A&A...468L..49M,2015ApJ...799...26M}. Rather than a continuous flow, the nuclear outflow in M51 appears to be composed of discrete clumps. However, since we cannot resolve those substructures even with our highest resolution of $\sim 1''$ (40\,pc), we calculate the outflow rates relying on the continuous conical model. This is equivalent to obtaining a representative outflow rate averaged over spatial scales of $\sim 1''$ (40\,pc), and, therefore, averaged over timescales of $\sim 10^5$\,yr. 
}

We estimate the outflowing gas mass based on the molecular emission that shows clear departures from the disc kinematics in Fig.\,\ref{fig:pv}; we add up the emission from the central $r \lesssim 2.5''$ over channels associated with outflow velocities (more than 30\,km/s apart from the systemic velocity). We assume a conversion factor $\alpha_\mathrm{CO} \sim \frac{1}{2} \times \alpha_\mathrm{CO, MW} = 2.2\,M_\odot$\,(K\,km\,s$^{-1}$\,pc$^2)^{-1}$, following the careful LVG analysis from \citet{2007A&A...468L..49M} for the centre of M51. We assume a similar reduction relative to the Galactic value for the factor that converts HCN luminosity to dense gas mass, $\alpha_\mathrm{HCN} \sim \frac{1}{2} \times \alpha_\mathrm{HCN, MW} = 5\,M_\odot$\,(K\,km\,s$^{-1}$\,pc$^2)^{-1}$, based on the Milky Way measurement from \citet{2004ApJS..152...63G}. The CO conversion factor is in agreement with \citet{2009A&A...493..525I,2009A&A...506..689I}, \citet{2013ApJ...764..117B} and \citet{2013ApJ...777....5S}, who found that $X_\mathrm{CO}$ can be up to a factor $2-10 \times$ lower than $X_\mathrm{CO}^\mathrm{MW}$ in the central $\sim$1\,kpc of a set of nearby galaxies with solar metallicity. The HCN conversion factor is compatible with the measurements from \citet{2012A&A...539A...8G} for other active galaxies, and the reduction of $\alpha_\mathrm{HCN}$ with respect to the Milky Way value is also necessary to prevent the dense gas mass from exceeding the total gas mass. This results in a total outflowing molecular gas mass of $M_{\mathrm{H}_2}=4.1 \times 10^6\,M_\odot$, and a dense gas mass of $M_\mathrm{dense}=2.7 \times 10^6\,M_\odot$.

We choose $R_\mathrm{out}\sim 1'' = 37$\,pc as the typical outflow radius (calculated as the luminosity-weighted average); $V_\mathrm{out} \sim 100$\,km/s is the characteristic velocity of the material that shows departures from the disc. The angle $\alpha$ \rev{that the outflow makes with the line of sight} is largely unconstrained, as the central area is mostly unresolved and we cannot determine the exact geometry. However, if we make the assumption that the outflow is parallel to the radio jet, we can constrain $\alpha = 70^\circ$ \citep[based on][as described in detail in Appendix\,\ref{angle}]{1988ApJ...329...38C}. This results in the following estimate for the outflow rate of bulk molecular gas, under the conservative assumption that it fills the cone:

\begin{equation}
\frac{\mathrm{d}M\,(\mathrm{H}_2)}{\mathrm{d}t}= 0.9\,M_\odot/\mathrm{yr},
\end{equation}
\\
and the following value for the outflow of dense gas:

\begin{equation}
\frac{\mathrm{d}M\,(\mathrm{dense})}{\mathrm{d}t}= 0.6\,M_\odot/\mathrm{yr}.
\end{equation}

This suggests that the molecular outflow is predominantly made up of dense gas. The values expressed as a function of $\alpha$ are $\dot{M}_{\mathrm{H}_2}=0.3\,M_\odot/\mathrm{yr} \times \tan (\alpha)$ and $\dot{M}_{\mathrm{dense}}=0.2\,M_\odot/\mathrm{yr} \times \tan (\alpha)$.
In Sect.\,\ref{Sec:discussion} we discuss how these values compare to those found for other nearby and distant molecular outflows.

The p-v diagrams that we have just presented show, overall, a lack of blueshifted counterpart to the observed redshifted velocities. 
There are different reasons that could explain this, but the most simple one is to assume that it is associated with asymmetric AGN activity (a hypothetical one-side activity cycle that would explain why the jet is stronger in the south). Indeed, the structure that \citet{2007A&A...468L..49M,2015ApJ...799...26M} resolve near the nucleus, and argue that it is undergoing an outflow, is elongated towards the south, where the jet is stronger, and corresponds to redshifted velocities.
In this dynamical picture, it is conceivable that, while only the redshifted component is seen now, a blueshifted counterpart might have existed in the past. We note that there are hints of a blueshifted component in HCN; examining the HCN profiles in the central area we have confirmed that this is real emission, and the fairly Gaussian HCN line shows a longer blue tail near the nucleus (this is also seen in the gradient in the velocity field of Fig.\,\ref{fig:GaussianMom}). However, higher angular resolution data is required to unambiguously address this issue.

\subsubsection{Dust extinction and nebular emission in the nucleus: the central ``bump''} 
\label{Sec:bump}

\begin{figure*}[t]
\begin{center}
\includegraphics[width=1.0\textwidth]{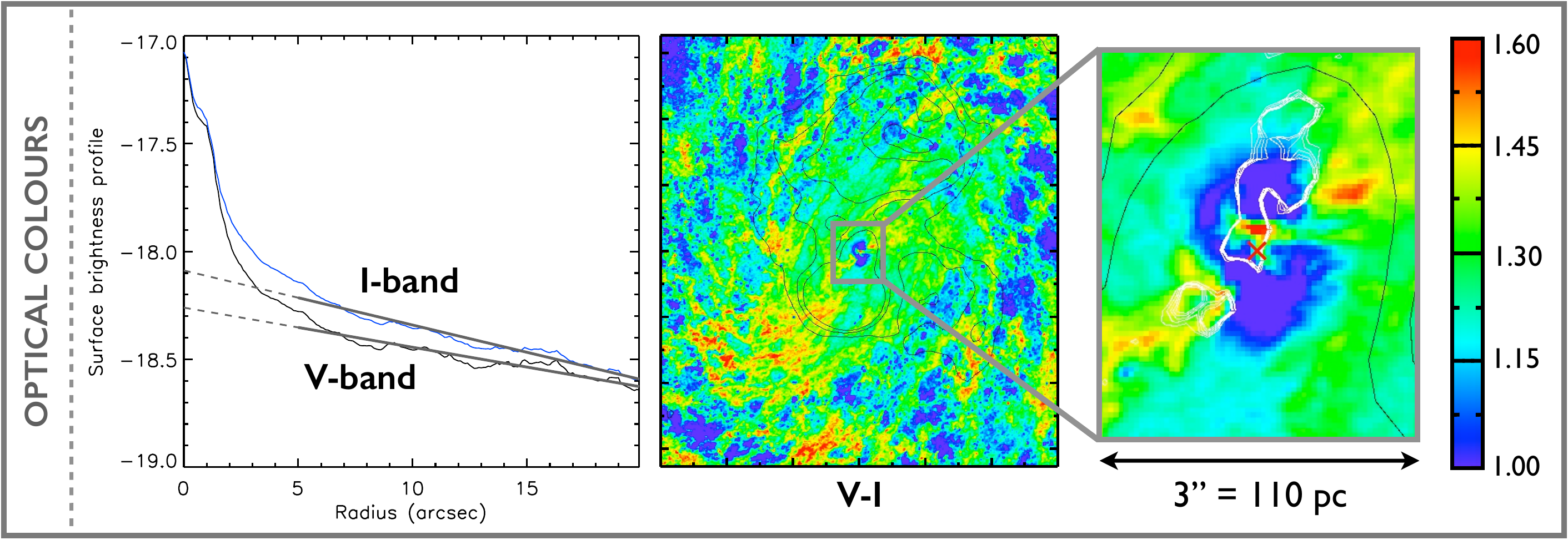}
\end{center}
\caption{\textit{From left to right:} radial profile of surface brightness in the HST $I$-band and $V$-band, in units of mag/arcsec$^2$; $V-I$ colour map of the same field of view shown before, $30'' \times 33''$, with large tickmarks indicating a separation of $5''$ (in linear scale from 1.0 to 1.6); blowup of the central area, where the ``bump'' is originating from (overlaid with H$\alpha$ contours in white; the black contours correspond to the radio jet traced by 20\,cm continuum).
}
\label{fig:colmap}
\end{figure*}

Figure\,\ref{fig:colmap} shows the radial surface brightness profile of the central $r<20''$ of M51 for different optical bands (HST $V$-band and $I$-band). The central $r \lesssim 5''$ cannot be fitted with the same S\'{e}rsic profile as the rest of the disc, and the central excess of light with respect to the S\'{e}rsic component is $\sim$1\,mag. We propose that this excess of light probably stems from a combination of two effects: (a) unobscured stellar continuum relative to the surroundings because dust has been largely evacuated in $r \lesssim 5''$; and (b) in the centremost $r \lesssim 2''$, nebular emission likely has a significant contribution in the optical bands, reflecting on a feature of distinct blue optical $B-V$ and $V-I$ colours (Fig.\,\ref{fig:colmap}). This structure is responsible for the bump that the optical surface brightness profiles show close to the nucleus.

The very blue optical colours ($B-V=0.3$, $V-I=0.9$) could be indicative of a young stellar population.
The excess of light was already attributed to a nuclear star cluster by \citet{1997AJ....113..225G}, who showed that an east-west dust lane is crossing the centre (the position of the excess light). Fitting a King profile they estimate an upper limit to the core radius of this alleged stellar cluster, $r_c = 0.3''$ (14\,pc). According to \citet{2011ApJ...740...42L}, the $V-I$ and $B-V$ colours can indeed be explained by a (very) young stellar population ($\lesssim 10$\,Myr, labelled ``P2'' in their paper); therefore, if the cluster hypothesis is true, this is likely a stellar cluster in the process of getting born. This is slightly contradictory with the fact that star formation tracers do not show evidence for a significant amount of recent star formation in the centre of M51 \citep{2007ApJ...671..333K,2015RAA....15..802F}, and also the fact that the centre shows [3.6]-[4.5] near-infrared colours characteristic of non-stellar emission (see \citealt{2015arXiv151003440Qalt}).
 Additionally, when inspected carefully, the shape of the colour maps is not consistent with the expectations for a relaxed stellar system (even though the partial obscuration due to the central dust lane and the finite resolution of the HST images can contribute to the observed appearance, at least partially).
 We have extracted a central spectrum from the VIRUS-P IFU dataset of M51 presented in \citet{2009ApJ...704..842B} and run Gandalf to identify the stellar populations that best explain the spectrum in the central IFU pointing (RA$=$13:29:52.673, Dec$=$+47:11:43.62). We find no significant evidence for young stars, with an overall luminosity weighted age of 8.9\,Gyr; however, the spatial resolution of this dataset ($5.5''$) is insufficient to conclusively rule out a small young stellar population in the very centre ($1''$). 

 Alternatively, the inner bump could be due to scattered light from the AGN \citep[e.g.][]{2016MNRAS.456.2861O}. If scattered light is the reason, one would in principle expect (bi)conical geometry, and not the arc-like structure that we see $\sim$1$''$ north of the AGN position; however, scattering cones often show significant asymmetries \citep{2016MNRAS.456.2861O}. This hypothesis was already suggested by \citet{2011ApJ...740...42L}.
  If indeed due to scattered light, the reason why we see this emission only in the north of the AGN, and with such an peculiar shape, could be because the radiation is scattered by a gas cloud of higher density which has that specific shape; this would also help explain why the ionisation region in the north is not as extended as in the south of the AGN (an important part of the ionising optical radiation would be scattered by this structure, preventing ionisation further north).
  Polarisation studies would be useful to confirm whether the excess optical emission in the centre of M51 is indeed due to scattered light from the AGN.
Probing the true nature of this structure is beyond the possibilities of our current data, but this intriguing structure reinforces the idea that the processes taking place near the nucleus of M51 are many and complex.

\section{Discussion}
\label{Sec:discussion}

\begin{figure*}[t]
\begin{center}
\includegraphics[width=1.0\textwidth]{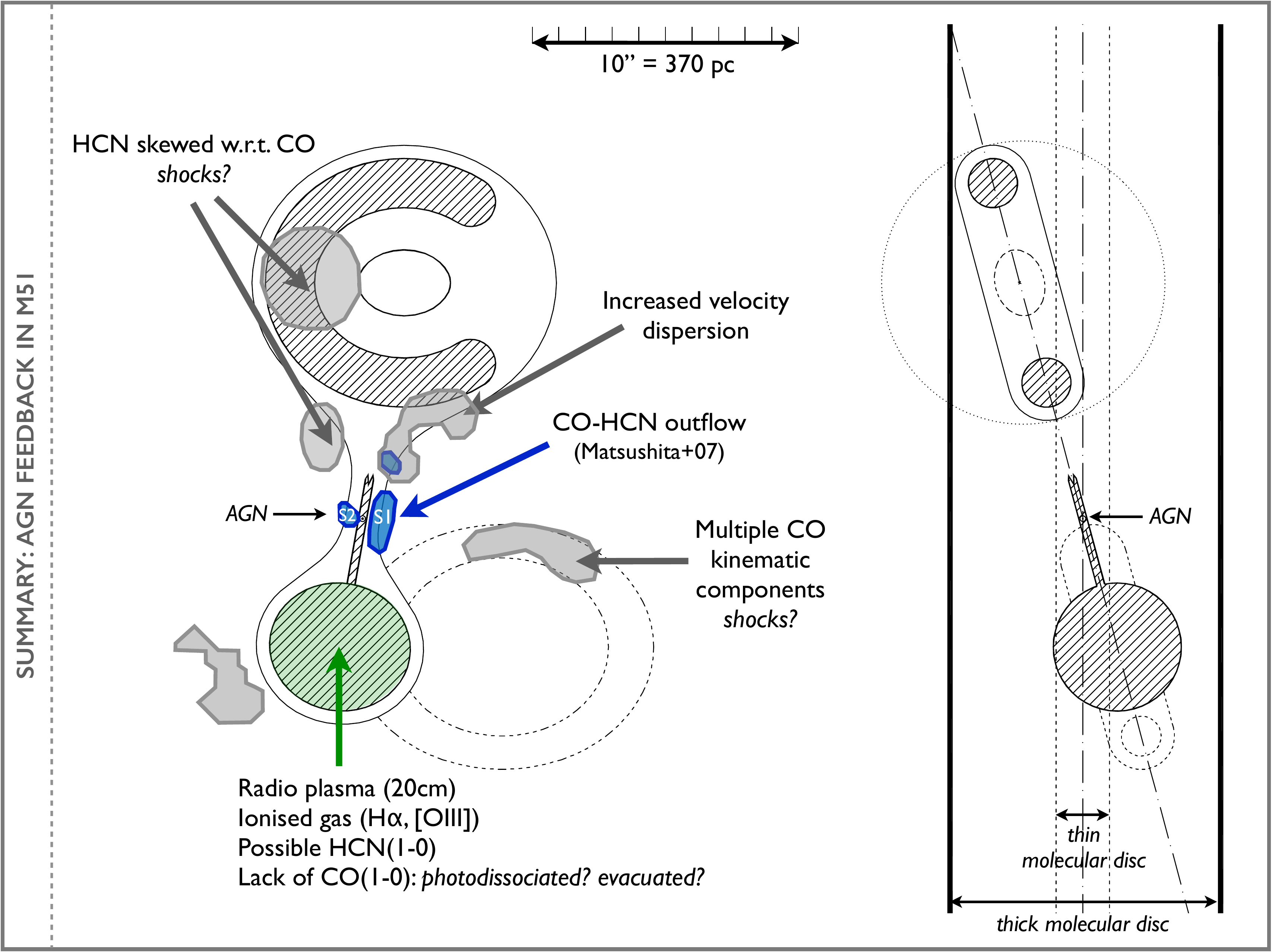}
\end{center}
\caption{Cartoon showing the scales relevant to the different processes involved in AGN feedback in the nucleus of M51. The striped areas delinate the main elements of the radio plasma jet (synchrotron emission); the nuclear collimated jet was resolved by \citet{1992AJ....103.1146C}, and the rest of the structures have been identified on the maps from \citet{2011AJ....141...41D}. The left panel corresponds to a face-on view, whereas the right panel shows a possible edge-on deprojection.
}
\label{fig:cartoon}
\end{figure*}

We have presented a number of results related to the active nucleus of M51. 
Overall, these pieces of information delineate a complex situation, in which processes that operate on different spatial scales are interconnected. In the central $\lesssim 3''$ (110\,pc) next to the AGN, molecular gas has been suggested to be entrained by the nuclear radio jet (Matsushita et al.~2007), which is collimated out to a distance of  $r \lesssim 2.3''$ (85\,pc) in the south ($r \lesssim 1.5''=60$\,pc in the north). Both in \mbox{CO(1-0)} and \mbox{HCN(1-0)} emission we find extremely broadened lines in this central area, preferentially redshifted ($\Delta v \sim 100$\,km/s, $\sigma \sim 150$\,km/s), and with \rev{decreasing CO/HCN} line ratios which approach unity near the centre. In the optical, a very bright and blue structure $\sim$1$''$ (37\,pc) north of the AGN becomes apparent through HST imaging, but its nature and potential interplay with the outflowing (or inflowing) gas remains elusive. At the scales spanned by the ionised optical emission lines (H$\alpha$, [NII], [OIII]), out to $\sim$1$''$ (37\,pc), we find a clear scarcity of CO emission ($\times 10$ lower flux), which we probe down to $1''$ scales; the ionisation cone seems to be associated with a dearth of bulk molecular gas, whereas clouds of CO accumulate towards the edges of this ionisation cone. At larger spatial scales, including the northern radio-emitting loop, and the boundaries of the southern XNC, we find evidence of the radio jet impacting the surrounding molecular material. This manifests itself in the form of multiple kinematic components, a \rev{reduced CO/HCN} ratio, and strong deviations between the HCN and CO line profiles; the HCN line is typically broader and more skewed (even at matched resolution). 
We recall that in this paper we refer by \textit{radio jet} to all the structures involved in the presumably expanding plasma, which produces synchrotron emission, and are detected at radio wavelengths (3.6, 6, 20\,cm); the collimated radio jet itself is $\sim$4$''$ (150\,pc) long, and $\sim$0.3$''$ (12\,pc) wide: it connects to the XNC in the south, and points towards the ``C''-shaped loop in the north (but is not directly connected to it). The cartoon in Fig.\,\ref{fig:cartoon} shows the different components that we will discuss and their relative sizes, in an attempt to illustrate the geometry and the scales involved.

\subsection{Under-luminous CO in the ionisation cone}

There are several possibilities to explain the scarcity of CO emission in the area covered by the jet relative to the edges. First, CO gas could simply be photodissociated by the strong radiation from the non-stellar nuclear source, the AGN, as it enters the ionisation cone; this seems plausible, as energetic arguments also point to the radiation field from the AGN as the most likely source of ionisation leading to the H$\alpha$, [OIII], and [NII] optical emission (Bradley~et~al.~2004). Another possibility that is compatible with our observations is that the CO-emitting clouds have been mechanically evacuated by the expanding radio plasma jet. The fact that the southern side is presently more active (the collimated radio jet is longer towards the south, brighter, and more straight) is probably the consequence of an alternating one-side activity cycle \citep{2015MNRAS.452...32R}. It is also possible that 
 radiative transfer mechanisms are leading to reduced \mbox{CO(1-0)} emission in that ionised area or, conversely, radiative transfer effects that result in enhanced emission at the edge of the cone \rev{\citep[this enhanced emission could also be the result of radiative shocks associated with the radio jet;][]{2009A&A...502..515G}}. Having said that, the masing effects identified by \citet{2015ApJ...799...26M} are only expected to be relevant sufficiently close to the AGN \citep{2014A&A...567A.125G}.
In any case, all three hypotheses (\rev{photodissociation}, mechanical evacuation, or radiative transfer effects) would explain why there is more ionised optical emission in the south, and also why the scarcity of \mbox{CO(1-0)} is more severe on that side. Finally, it is also possible that the magnetic field created by the expanding, probably relativistic jet plays a role in accumulating molecular clouds towards its edges (e.g.~by ambipolar diffusion, as proposed by Krause et al.~2007 for NGC\,4258).

The sharp picture that PAWS provides at $1''$ (Fig.\,\ref{fig:COhole}) rules out a possible chance alignment of ionised gas, radio jet, and lack of CO emission. Of course, in other galaxies projection effects could conceal such an effect, even if the ionisation cone or jet has indeed evacuated or photodissociated the bulk molecular gas: if the cone is narrow enough, or if it is more inclined than in M51, it can well be that  molecular clouds are detected flowing in front or behind the ionisation cone, and being unable to tell their relative height in the disc when seen in projection. M51 offers the unique setting of being seen almost face-on, with the jet almost exactly coplanar with the disc, and a wide opening-angle ionisation cone \citep[$74^\circ$; ][]{2004ApJ...603..463B} which blocks most of the molecular (thin) disc, and it is thus ideal to test the relative distribution of molecular gas, ionisation cone, and radio jet. What we have shown is that, at least in the area with strong ionised optical line emission, which overlaps with the radio jet, M51 is largely depleted of \mbox{CO(1-0)} emission. 

Naturally, from our result on a single galaxy, one cannot claim that this is a typical behaviour: the current situation in  M51 can certainly be special. However, the other few cases where it was possible to spatially resolve the interplay between radio jets and molecular gas also point in the same direction.
Krause et al.~(2007) showed that in NGC\,4258 a funnel almost completely devoid of \mbox{CO(1-0)} exists along the radio jet, with molecular gas accumulating at the edge of the jet in two parallel CO ridges  (2.8\,kpc in length, and separated by 175\,pc from each other). Similarly to M51, the jet has a low ($15-30^\circ$) inclination to the disc.
Interestingly, in NGC\,4258, ionised gas emission traced by H$\alpha$ also peaks in the centre of the funnel, where CO is lacking; the same situation as we find in M51. From their careful analysis of the position-velocity diagram of IC\,5063 obtained with ALMA, Morganti et al.~(2015) propose a scenario which would also go in the same direction. They detect signatures of lateral expansion in this edge-on galaxy, and through comparison with numerical simulations by \citet{2011ApJ...728...29W} and \citet{2012ApJ...757..136W}, they suggest that a radio jet expanding through a porous medium is responsible for their observations; as it expands, the jet would look for paths of least resistance, and clear molecular gas from its way. Finally, \citet{2014A&A...567A.125G}, who study NGC\,1068 with ALMA using multiple molecular transitions, also find a ``hole'' in the molecular emission next to the AGN, which could be indicative of gas having been evacuated. 

It would be extremely interesting to confirm whether HCN emission also shows the same behaviour and equally avoids the ionisation cone in M51. Unfortunately, we cannot conclusively prove this with our current $3-4''$ HCN dataset (Fig.\,\ref{fig:Resol}). 
In fact, we find similar HCN flux levels between the ionisation cone and surroundings as we do for CO when matched to $4''$ resolution. Therefore, it is possible that a similar spatial distribution applies to the dense gas, which would not be surprising because both \rev{photodissociation} and mechanical evacuation are in principle expected to affect both phases of the molecular gas similarly.
However, higher-resolution observations of the whole ionisation cone in HCN are required to confirm this conjecture.

\subsection{HCN-CO differences along the jet}
\label{HCN-COdiff}

We have evaluated the difference between HCN and CO profiles, after scaling them to match the same peak brightness line temperature, as a diagnostic tool to identify potential regions in which molecular gas is impacted by the radio jet. The agreement of this HCN-CO offset emission with the area covered by the radio jet is very good, and we interpret this as direct evidence that the AGN is impacting the molecular gas disc through the radio plasma jet; at the same time, the fact that this effect becomes visible when looking at the difference between CO and HCN profile shapes implies that this is a \textit{differential effect}, with enhanced HCN excitation which could reveal a larger impact of AGN feedback through radio jets on the dense gas than on the bulk molecular gas. 

Kinematically, the situation highlighted by the HCN-CO velocity map is complex, with a quickly 
varying net velocity offset between CO and HCN. This can be the result of intrinsically rapid positional variations in the differential response imprinted on the gas traced by HCN and CO; alternatively, perhaps a relatively smooth variation results in the observed map due to projection  and beam dilution effects.
New observations at matched $1''$ are required to disentangle beam dilution from other effects and build a sharper picture of the feedback kinematics in the different phases of molecular gas in M51.

But not only is the response of HCN different from CO; when looking at the two tracers individually, we also find distinct kinematic components in their line profiles for some regions.
These differences have motivated our attempt to automatically separate the emission into multiple Gaussian contributions. The first (main) Gaussian component reflects the intensity distribution and velocity field expected for a differentially rotating disc; the second component, however, shows a velocity gradient from north-east (blueshifted) to south-west (redshifted), probably the result of galactic rotation  of shocked gas in combination with the outflow.
There are some intrinsic difficulties in performing Gaussian separations on a pixel-by-pixel basis, especially when the properties of the second components change quickly with position (as in M51), and do not have a very high signal-to-noise ratio. 
However, the mere presence of these secondary components, and the fact that they are so susceptible to local changes, is already telling us something important: similarly to the effects that we identified when analysing the differential HCN-CO kinematic behaviour, the analysis of the CO and HCN profile shapes independently points to a complex feedback situation, compatible with the jet pushing laterally (which due to projection leads to quickly changing red- or blueshifted components).
It is also important to emphasise that both effects are not always redundant: in some positions, CO and HCN show important relative differences, while none of them has a second Gaussian component (HCN is typically broader or skewed). In other positions, we see multiple components, but no difference between the HCN and CO lines: they are scaled replicas, and both have multiple components which coincide in velocity and shape. 

The radio jet is far from being a smooth structure when inspected closely, showing kinks, asymmetries, and irregularities; the molecular gas distribution is also far from smooth, and rather clumpy \citep{2013AJ....146...19L,2014ApJ...784....3C}. Therefore, if the expanding jet is pushing on clumpy gas, we can expect quickly varying kinematic components, especially if the imprinted velocities are essentially coplanar within the galaxy.
In other words, the complex kinematic response in the molecular gas might reflect the underlying non-smooth distribution of the ISM (and the geometrical irregularities of the radio jet itself).

Altogether, this is clear evidence that the molecular gas is impacted by the radio jet, out to radii of $\sim$500\,pc, even in a system with an AGN as weak as the one in M51. 

\subsection{Consequences of AGN feedback}

One of the most important consequences of AGN feedback as we see it in M51 is that it will contribute to inject turbulence in the molecular disc, which has the potential of making the gas unable to form new stars. With our new observations, we have found that velocity dispersion is enhanced at least out to $r \sim 5''$, \textit{further out than the nuclear outflowing structure identified by Matsushita et al.~(2007)}.
 It is worthwhile noting that what we resolve as multiple Gaussian components would be attributed to mere turbulence when observed at lower resolution (e.g.~from single-dish data, or from interferometric observations with lower resolution); therefore, we speculate that part of the increased velocity dispersion measured in other galaxies at lower spatial resolution could ultimately be resolved into independent components which vary rapidly from position to position. Increased turbulence from AGN feedback is now recognised as an agent which can regulate star formation, with quantitative measurements of such star formation suppression existing for some early-type galaxies such as NGC\,1266 \citep{2011ApJ...735...88A,2013ApJ...779..173N,2014ApJ...780..186A}. \rev{Our results also agree with the picture recently proposed by \citet{2015A&A...574A..32G}, in which AGN feedback can suppress star formation through a combination of molecular outflows and the dissipation of a small ($< 10\,\%$) fraction of the kinetic energy of the radio jet under the form of supersonic turbulence.}

The increase of turbulence and the appearance of strong kinematic effects as jets propagate through a molecular medium are expected on theoretical grounds, especially at the interaction points between jets and dense gas clouds \citep[e.g.][]{1992pagn.conf..307W}. Here we find that the feedback is spatially extended, and does not only involve the collimated nuclear jet, but also the large-scale plasma structures. We also find that HCN is more affected than CO, perhaps due to different excitation conditions (enhanced radiative transfer effects triggered by shocks). \rev{It is worth mentioning that galactic rotation will provide an additional term of pressure against the propagation of the radio jets \citep{2013ApJ...779...45M};
although we expect the jet lifetime to be shorter than the rotation timescale (see Sect.\,\ref{Sec:jet}), modelling this is detail is currently difficult given the lack of measurements of the jet propagation speed.}

Observations of more molecular species could help confine the precise role of shocks. The flux measurements of the different CO transitions (the ``CO ladder'') from \textit{Herschel}/SPIRE fits by \citet{2015PhDT.........1S} for the nucleus of M51 show a characteristic flattening at high-J values; an analogous situation has been found in Mrk\,231 and NGC\,6240, and explained using photon-dominated regions (PDRs), X-ray-dominated regions (XDRs), and shocks \citep{2010A&A...518L..42V,2013ApJ...762L..16M}. This suggests that, probably, both XDR and shocks are important in the nucleus of M51; to perform a more quantitative analysis in this sense, however, observations of additional molecular transitions will be essential, ideally \rev{including observations of warm H$_2$ and cooling lines such as [CII]} at sufficiently high spatial resolution to probe the different regions that we have identified in this paper. \rev{This would allow to calculate a more complete energy balance, and it would permit to further quantify the amount of kinetic energy dissipated through shocks and turbulence (as well as the associated timescales).}

\subsection{Nuclear outflow}

As we have already pointed out, there is a strong bimodality between the line profiles (both CO and HCN) in the central $\sim$5$''$ and the rest of our field of view. In the central area, it becomes essentially impossible to isolate different Gaussian contributions, and all we find is a very broad approximately Gaussian line, which is preferentially redshifted. This explains the strong flux at redshifted velocities in the p-v diagrams of Fig.\,\ref{fig:pv}. In HCN, a small blueshifted counterpart also becomes clear, symmetrical with respect to the centre. When looking at the variation of velocity with position within this area, we also find a noticeable velocity gradient in the northeast-southwest direction (from blue- to redshifted), especially clearly traced by HCN (Fig.\,\ref{fig:GaussianMom}). This is compatible with the velocity gradients found by \citet{2007A&A...468L..49M,2015ApJ...799...26M}, who attribute them to a molecular gas outflow; the emission they observe is coming from a number of resolved clumps. Our continuous gradient could be the result of an intrinsically smooth velocity variation (with part of the flux missed by the Matsushita et al. interferometer-only observations, see below), or because we are smoothing the structures that Matsushita et al.~identify, producing a continuous appearance.

Even though Matsushita et al.~(2015) do not quote the integrated fluxes for \mbox{CO(1-0)} and \mbox{HCN(1-0)}, we can estimate them from their Fig.\,2 and Fig.\,6 to be $\sim$10\,Jy\,km/s and $\sim$5\,Jy\,km/s, respectively, in the inner $r<3''$. With our maps, which include short-spacing corrections and therefore recover all the flux, we measure 25.0\,Jy\,km/s in \mbox{CO(1-0)} and 7.2\,Jy\,km/s in \mbox{HCN(1-0)} for the same region ($r<3''$); therefore, Matsushita et al.~seem to be missing $\sim$50\% of the flux in CO and $\sim$30\% of the flux in HCN for the central area. Comparing our HCN map with and without the short spacing correction out to $r<20''$, we confirm that the interferometer recovers 43\% of the total HCN flux from the 30m single-dish telescope; in CO, 37\% of the flux is recovered by the interferometer at $1''$ resolution, and about 50\% at $3-6''$ resolution \citep{2013ApJ...779...43P}.
This means that, while part of the differences observed between Matsushita et al.~and this work could stem from the different resolutions achieved, part could also be due to the lack of flux coming from diffuse structures, which gets filtered by the interferometer.

Now, we briefly discuss the geometry of the outflow identified by \citet{2007A&A...468L..49M}, and how it fits in the global picture of AGN feedback that emerges from our work. The overlay with the radio jet map shown by \citet{2015PKAS...30..439M} makes it clear that the structures S1 and S2 \citep[first identified by ][]{1998ApJ...493L..63S} lie at opposite sides of the radio jet (next to the estimated position of the AGN), whereas S3 lies $1.5''$ towards the north, precisely
where the putative counter-jet seems to dissolve (see Fig.\,\ref{fig:cartoon}). 
Even though \citet{2007A&A...468L..49M} and \citet{2015PKAS...30..439M} claim good agreement between the molecular outflow and the ionised outflow, this is in tension with the ionised gas observations: Bradley et al.~(2004) assume that the entrained ionised component is Cloud 1, which is blueshifted in the south, as expected from the geometry for a radial outflow if we assume that the southern cone is closer to us (Bradley et al.~2004). However, the gradient that Matsushita et al.~measure goes exactly in the opposite direction, with redshifted velocities \textit{in the south}. This apparently contradicting result can be reconciled if we reverse the assumed geometry of the jet, so that the southern cone is pointing away from us, as we briefly commented in Sect.\,\ref{Sec:distribution} and Sect.\,\ref{HCN-COdiff}. This possibility would also explain the entrainment of a higher amount of the ionised clouds identified by  Bradley et al.~(2004); instead of only Cloud 1 being compatible with entrainment along the jet, at least Clouds 4, 4a, and 3 would be. We note that the HCN blueshifted counterpart identified in the p-v diagram on Fig.\,\ref{fig:pv}, which lies $\sim$1$''$ at the west of the nucleus (as opposed to the prominent redshifted material at the east), might be indicative of a dense molecular outflow which is \textit{perpendicular to the jet}, propagating in the plane of the galaxy. This idea would be supported by the velocity gradient from north-west to south-east that we found in Fig.\,\ref{fig:GaussianMom} (if true, we would be resolving an outflow perpendicular to the line-of-nodes).

With our observations, we can also estimate the kinetic energy and momentum of the molecular gas that is presumably outflowing. These have already been estimated by \citet{2004ApJ...616L..55M,2007A&A...468L..49M}; however, our estimates should be more accurate, as they include short spacings corrections, and therefore recover all the flux.

The kinetic luminosity is given by:

\begin{equation}
L_\mathrm{kin}=\frac{1}{2} \times \frac{\mathrm{d}M}{\mathrm{d}t} \times \left(\frac{V_\mathrm{out}}{\cos (\alpha)}\right)^2,
\end{equation}

\noindent
and the momentum flux can be obtained as:

\begin{equation}
\frac{\mathrm{d}P}{\mathrm{d}t}= \frac{\mathrm{d}M}{\mathrm{d}t} \times \frac{V_\mathrm{out}}{\cos (\alpha)}.
\end{equation}

\noindent
Following the conservative assumptions made in Sect.\,\ref{Sec:pvdiagrams} (multi-conical outflow uniformly filled by molecular gas), for $\dot{M}_{\mathrm{H}_2}=0.3\,M_\odot/\mathrm{yr} \times \tan (\alpha)$ and $\dot{M}_{\mathrm{dense}}=0.2\,M_\odot/\mathrm{yr} \times \tan (\alpha)$, taking again $V_\mathrm{out} = 100$\,km/s as the characteristic velocity, for CO we obtain  $L_\mathrm{kin}=2.35 \times 10^{40}$\,erg/s ($1.5 \times 10^{40}$\,erg/s for HCN), $\mathrm{d}P/\mathrm{d}t = 1.6 \times 10^{33}$\,g\,cm\,s$^{-2}$ (for HCN, $1.0 \times 10^{33}$\,g\,cm\,s$^{-2}$). \citet{2007A&A...468L..49M} provide the related quantities kinetic energy and momentum instead; our equivalent measurements would be: 
$E_\mathrm{kin}=3 \times 10^{54}$\,erg, $P = 2.4 \times 10^{47}$\,g\,cm\,s$^{-1}$, whereas the results from   \citet{2007A&A...468L..49M} were $E_\mathrm{kin}=3 \times 10^{52}$\,erg, $P = 8 \times 10^{45}$\,g\,cm\,s$^{-1}$. Therefore, the differences are very significant (our results being about two orders of magnitude higher). These values are still an order of magnitude lower than the energetics involved in NGC\,1068, for example \citep[$L_\mathrm{kin}=5 \times 10^{41}$\,erg/s, $\mathrm{d}P/\mathrm{d}t = 6 \times 10^{34}$\,g\,cm\,s$^{-2}$;][]{2014A&A...567A.125G}.

The star formation rate in the centre of M51 is very low, $\mathrm{SFR} (r<3'') \sim 0.01\,M_\odot$/yr in the central area involved in the putative outflow \citep[][]{2007ApJ...671..333K,2015RAA....15..802F}. Given that the star formation rate is two orders of magnitude lower than the outflow rate, it is in principle not possible that the outflow is driven by a nuclear starburst. The bolometric luminosity of the AGN is $L_\mathrm{bol} \sim 10^{44}$\,erg\,s$^{-1}$ \citep{2002ApJ...579..530W}; therefore, on energetic grounds, it is well possible that the AGN is driving the outflow. The question is how: in principle, gas could be expelled directly by radiation pressure, but as we have seen, CO emission seems to avoid the ionisation cone (which also seems to apply to the central $\sim$3$''$, judging from our moment-0 map, and the higher resolution maps from Matsushita et al.). Consequently, it looks like entrainment through the radio jet is the most probable mechanism. However, while the energy of the jet is estimated to be $6.9\times 10^{51}$\,erg, the momentum is only $2\times 10^{41}$\,g\,cm\,s$^{-1}$ (assuming a jet velocity of $0.9c$; Crane \& van der Hulst 1992). As already discussed by \citet{2004ApJ...616L..55M,2007A&A...468L..49M}, the energy is sufficient, but the power of the jet is too low to explain this coupling directly; perhaps a continuous transfer between energy and momentum could explain this apparent mismatch. \rev{We also note that the energy and momentum of the jet could in principle be underestimated, because the emissivity of the relativistic plasma depends on the local conditions of the ISM, and might not always be detectable in the radio maps.}. Finally, another possibility would be to abandon the idea of a molecular outflow, and interpret the velocity gradient observed as evidence of molecular gas \textit{inflow}. 

\rev{Finally, we briefly comment on the potential impact of this outflow of molecular gas on its host galaxy. If the material is outflowing, and if it shares the same inclination as the radio jet ($\alpha = 70^\circ$), this means that the projected velocities of $\sim$100\,km/s would translate into typical linear velocities around $\sim$300\,km/s. This is probably below the escape velocity for the whole galaxy, but sufficient to exceed the escape velocity of the bulge (which we estimate as $v_\mathrm{esc}^\mathrm{bulge} = \sqrt{2GM_\star / 5R_e} \sim 160$\,km/s).
Therefore, it can well be that a significant part of the molecular gas in the circumnuclear region is expelled from the bulge by the AGN-driven outflow, and the rest will be partially recirculated in a process similar to a galactic fountain, which could also increase the amount of turbulence and thus regulate star formation in the bulge region.
}

\subsection{Relation between inflow and AGN feedback}
In \citet{2015arXiv151003440Qalt} we have shown that there is molecular gas inflow down to our resolution limit of $1.7''$ (although the inner region, $r \lesssim 3''$ is highly uncertain). This molecular gas inflow can be explained by the gravitational torques exerted by the $\sim$1.5\,kpc-long stellar bar. However, in this paper we have shown that in the innermost region of M51 the interplay between AGN, molecular gas and stars becomes very complex. The apparently chaotic response of molecular gas to AGN feedback through radio jets could actually explain the discrepancies between AGN activity and large-scale inflow \citep[for example, the lack of correlation, or only weak correlation, between presence of bars and AGN activity;][]{2000ApJ...529...93K,2002ApJ...567...97L,2013ApJ...776...50C}. Overall, this would contribute to limit the amount of gas that can reach the SMBH, and, thus, control the AGN duty cycles. A scenario in which gas can easily make it to the central $\sim$100\,pc through secular evolution mechanisms, but is then trapped in a number of cyclical motions triggered by the omni-directional pressure from the radio jets is well compatible with our data.
It is quite remarkable that the inflow rate estimated in \citet{2015arXiv151003440Qalt} in the central $\sim$3$''$ coincides so well with the outflow rate estimated here, $\sim$1\,$M_\odot$/yr. This could be coincidence, of course, but it could also be reflecting some self-regulating balance between inflow and outflow.  \rev{Overall, the feedback effects that we have discussed (the molecular outflow, the lateral expansion, and the increase of turbulence) are a plausible way for the nuclear black hole to regulate its own growth in connection with its surrounding environment.}

It is worth briefly discussing the uncertainties involved in the outflow rates that we have calculated here. First and foremost, these rates rely on a strong methodological assumption, namely that all the molecular gas observed to have peculiar velocities near the centre is flowing out, and that it is doing so by uniformly filling a multi-conical volume. Judging from what we have seen in Sect.\,\ref{Sec:ionised}, this hypothesis is highly questionable for M51 (as optical ionised emission does fill an approximate bicone, but CO emission seems to accumulate towards its edges). Additionally, a number of systematic uncertainties are inevitably part of the estimation of outflow rates: the uncertain $X_\mathrm{CO}$ (which could easily vary by a factor of $\sim$2), the range of velocities selected, and the characteristic radius of the outflow, which cannot be easily determined. This means that, all in all, the outflow rate should only be regarded as an approximation to the order of magnitude.

The outflow rate in NGC\,1068 \citep[$\sim$60\,$M_\odot$/yr;][]{2014A&A...567A.125G}
is an order of magnitude higher than our estimation for M51 ($\sim$1\,$M_\odot$/yr); interestingly, the AGN bolometric luminosity is also one order of magnitude higher. The outflow rate in M51 is closer to that found by \citet{2013A&A...558A.124C} in NGC\,1433, $\sim$7\,$M_\odot$/yr (for a $L_\mathrm{bol} \sim 10^{43}$\,erg\,s$^{-1}$); the datapoint for M51 would appear as a lower outlier in the outflow rates--bolometric luminosity compilation from \citet{2015A&A...580A..35G}.

Of course, for (much) more active galaxies, such as those observed by \citet{2014A&A...562A..21C} at intermediate and high redshifts, the estimated outflow rates can be several orders of magnitude higher than what we find in M51. The question that immediately arises is whether the manifestations of AGN feedback in M51 are intrinsically different from those at play in very active galaxies, or if they are to some extent scaled versions. Specifically, we wonder if our finding that molecular gas seems to be largely depleted in the ionisation cone of M51 is also applicable to active galaxies undergoing powerful outflows. If CO is \rev{photodissociated} in M51, with such a low-luminosity AGN, it seems surprising that CO could survive in galaxies with even more intense nuclear radiation fields. 
In any case, it would be necessary to spatially resolve more spectacular outflows (with the handicap that they tend to be much more distant) to confirm the tantalising evidence provided by M51 and other nearby galaxies.

\section{Summary and conclusions}
\label{Sec:conclusions}

We have studied AGN feedback effects in a nearby spiral galaxy, M51, which hosts a low-luminosity active nucleus ($L_\mathrm{bol} \sim 10^{44}$\,erg\,s$^{-1}$) and a kpc-scale radio jet. The first important conclusion is that \textit{even with such a modest AGN}, the effects of feedback can be significant out to a distance of $\sim$500\,pc.

The particular spatial configuration of M51 has allowed us to directly witness the interplay between the radio jet and molecular gas, because the galaxy is almost face-on and its radio jet is expanding through the disc, at least in the inner 1\,kpc (the jet has an inclination $\sim$15$^\circ$ with the plane of the galaxy). The area of the jet where optical ionised lines are detected, the ionisation cone, is largely depleted of molecular gas as traced by \mbox{CO(1-0)} at $1''$ resolution. Instead, CO emission seems to accumulate \textit{towards the edges} of the ionisation cone. This is an important result, as it indicates that molecular gas may not survive under the strong radiation field produced by the AGN, and questions the applicability of (bi)conical outflow models to more distant, unresolved molecular outflows.

We find evidence for multiple components and disturbed kinematics in the molecular gas across the whole extent of the radio plasma jet. This becomes particularly clear when looking at the different kinematic response shown by CO and HCN, tracers of the bulk and dense phases of molecular gas, respectively. Therefore, relative differences between CO and HCN prove to be a useful diagnostic tool when it comes to probing feedback effects from radio jets. Mechanical shocks are the most likely explanation for the observed differences between both tracers. We also find increased turbulence (higher velocity dispersion) in the molecular gas across the whole region covered by the radio jet.

The situation found in M51 is analogous to that recently observed in other nearby galaxies with similarly modest radio jets (e.g.~Krause~et~al.~2007; Morganti~et~al.~2015), and agrees with numerical simulations of radio jets expanding through a clumpy medium \citep{2011ApJ...728...29W,2012ApJ...757..136W}.
Therefore, a new paradigm seems to be emerging, in which feedback through radio jets has complex implications for molecular gas (and, therefore, for star formation), probably pushing it in different directions and increasing its turbulence. It seems that outflows are not the only important consequences of AGN feedback; in addition to potential removal of molecular gas, injecting turbulence and therefore preventing molecular gas from forming new stars is an important part of the AGN response.

Overall, we have shown that the feedback from the AGN in M51 is a multi-scale phenomenon. In addition to the large-scale impact on molecular gas across the radio jet area, the central $5''$ (180\,pc) display a more extreme version of feedback, which has been interpreted as a molecular outflow before \citep{2007A&A...468L..49M}. We have estimated the corresponding outflow rates with our data, $\dot{M}_{\mathrm{H}_2} \sim 0.9\,M_\odot/\mathrm{yr}$ and $\dot{M}_{\mathrm{dense}} \sim 0.6\,M_\odot/\mathrm{yr}$, and discussed geometrical caveats. 
It is worth noting that the typical velocities of this \rev{outflow are below the escape velocity for the whole galaxy, but they are sufficient to bring a substantial amount of molecular material out of the bulge. Therefore, in combination with the increased turbulence induced by the radio jet, the feedback from the AGN could be sufficient to prevent M51 from growing a massive, young bulge. This would explain the low star formation rates within the bulge, and it would contribute to maintaining the approximately constant relation between the mass of the bulge and that of the central supermassive black hole. We could be uncovering the mechanisms that permit black holes to regulate themselves in concert with the growth of their surroundings.
}

It would be important to confirm whether similar multi-scale mechanisms operate in more active galaxies, which do indeed have significant amounts of gas at velocities large enough to escape their host. However, this will prove to be observationally challenging, as those very active sources tend to be more distant, and therefore much longer integration times are required  to achieve the same sensitivity with current-day interferometers; ALMA and NOEMA should be able to undoubtedly contribute in this direction.

One of the important conclusions from our study is that both high spatial \textit{and spectral} resolution are necessary to obtain a complete picture of feedback effects, as multiple velocity components and kinematic differences between tracers can only be robustly characterised when sufficient velocity resolution is available. In a similar way, we have confirmed that the response of tracers of gas \rev{with different critical densities} (CO, HCN) are not redundant, and provide important hints as to what regions are impacted by the radio plasma jets. Therefore, future observations should go in the direction of high-resolution, multi-species observations of active nuclei, probing sufficiently large spatial regions to cover the various spatial effects involved.

\small  
%
\begin{acknowledgements}   
Based on observations carried out with the IRAM Interferometer NOEMA. IRAM is supported by INSU/CNRS (France), MPG (Germany) and IGN (Spain).
\rev{The authors would like to thank the anonymous referee for very constructive comments.}
We would also like to thank Larry Bradley for kindly providing the data on the ionised outflow in M51 \citep[published in ][]{2004ApJ...603..463B}, and Mark Norris, Brent Groves, and Diederik Kruijssen for helpful suggestions.
We acknowledge financial support to the DAGAL network from the People Programme (Marie Curie Actions) of the European Union's Seventh Framework Programme FP7/2007- 2013/ under REA grant agreement number PITN-GA-2011-289313. M.Q.~acknowledges the International Max Planck Research School for Astronomy and Cosmic Physics at the University of Heidelberg (IMPRS-HD). M.Q.,~S.E.M.,~D.C.~and A.H.~acknowledge funding from the Deutsche Forschungsgemeinschaft (DFG) via grants SCHI~536/7-2, 
 SCHI~536/5-1, and SCHI~536/7-1 as part of the priority program SPP~1573 ``ISM-SPP: Physics of the Interstellar Medium.''
 S.G.B.~thanks support from Spanish grant AYA2012-32295. F.B.~acknowledges support from DFG grant BI 1546/1-1. G.B.~is supported by CONICYT/FONDECYT, Programa de Iniciaci\'{o}n, Folio 11150220. K.K.~acknowledges grant KR 4598/1-2 from the DFG Priority Program 1573. J.P.~acknowledges support from the CNRS programme ``Physique et Chimie du Milieu Interstellaire'' (PCMI).
\end{acknowledgements}

\bibliography{/Users/querejeta/Documents/E1_SCIENCE_UTIL/mq.bib}{}
\bibliographystyle{aa}{}

\appendix
\section{Kinematic model of the central region of M51}
\label{sec:TiRiFiC}

\begin{figure*}[t]
\begin{center}
\includegraphics[width=1.0\textwidth]{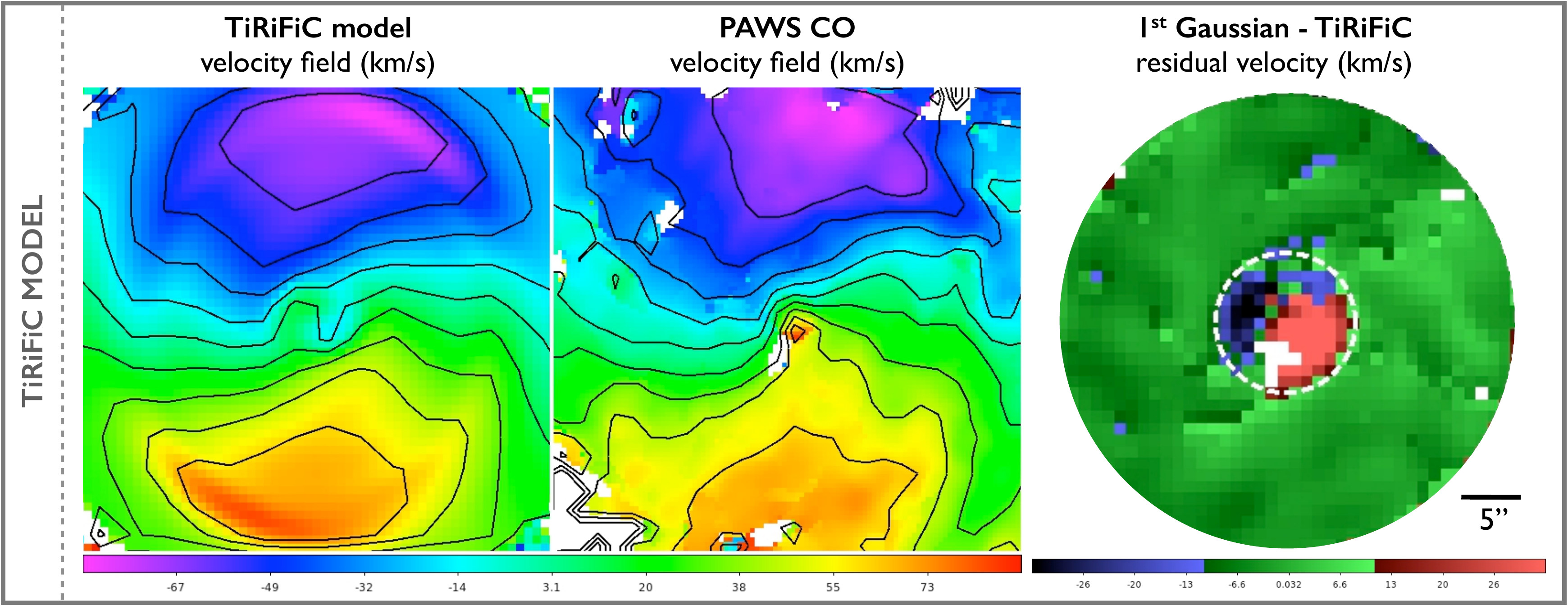}
\end{center}
\caption{Velocity field of our \texttt{TiRiFiC} model compared to the velocity field from PAWS (moment-1, at $3''$ resolution). The isovelocity contours correspond to the range [-60\,km/s, 60\,km/s] in increments of $\Delta v=15$\,km/s. The right panel shows the velocity difference between the first Gaussian fit (Sect.\,\ref{Sec:multiplecomp}) and the velocity centroid from our \texttt{TiRiFiC} model, highlighting their good agreement except in the central  $r \lesssim 5''$.
}
\label{fig:TiRiFiC}
\end{figure*}

Here we use a synthetic kinematic model of the \mbox{CO(1-0)} cube of M51 (PAWS at $3''$ resolution) to verify that the emission assigned to the first Gaussian does indeed correspond to the disc kinematic component. For that, we have constructed a \texttt{TiRiFiC} model \citep{2007A&A...468..731J} of M51, implementing a differentially rotating disc and an idealised $m=2$ bar \citep{2007ApJ...664..204S}. We start imposing the rotation curve of M51 parametrised by \citet{2013ApJ...779...45M}, and allow for the surface brightness and amplitudes of the $m=2$ bar
to be fitted in radial bins of $3''$, out to a radius of $r=22''$ (where the bar ends; \citealt{2010MNRAS.402.2462C}). Additionally, we allow for the PA and inclination of the disc to vary as a function of radius; this is because the tilted-ring strategy from \texttt{TiRiFiC} assumes a circular disc which projects into ellipses, but, as argued in \citet{2015arXiv151003440Qalt}, there seems to be an oval in the central region of M51, probably as the result of the interaction with NGC\,5195.
Fig.\,\ref{fig:TiRiFiC} shows the velocity centroid of the final \texttt{TiRiFiC} model, compared to the velocity centroid from PAWS (at $3''$ resolution).
We note that the \texttt{TiRiFiC} model becomes meaningless in the inner $\sim$5$''$, as the emission is dominated by the molecular outflow, and therefore the amplitudes of the bar in that innermost region cannot be constrained with our dataset; this area is marked with a white dashed circle on Fig\,\ref{fig:TiRiFiC}.

The \texttt{TiRiFiC} model confirms that the velocity centroid of our ``Disc Gaussian'' is in good agreement with the smooth disc-bar velocity model that we have constructed, as shown by the right panel of Fig.\,\ref{fig:TiRiFiC}. The velocity difference between Gaussian fit and model is typically 5\,km/s, and virtually always below 10\,km/s (the limits of the green region of the colourbar we use), except in the central $r \lesssim 5''$, as commented above.

\section{Maximum disc contribution in $r \lesssim 5''$}
\label{sec:disccontr}

\begin{figure}[t]
\begin{center}
\includegraphics[width=0.45\textwidth]{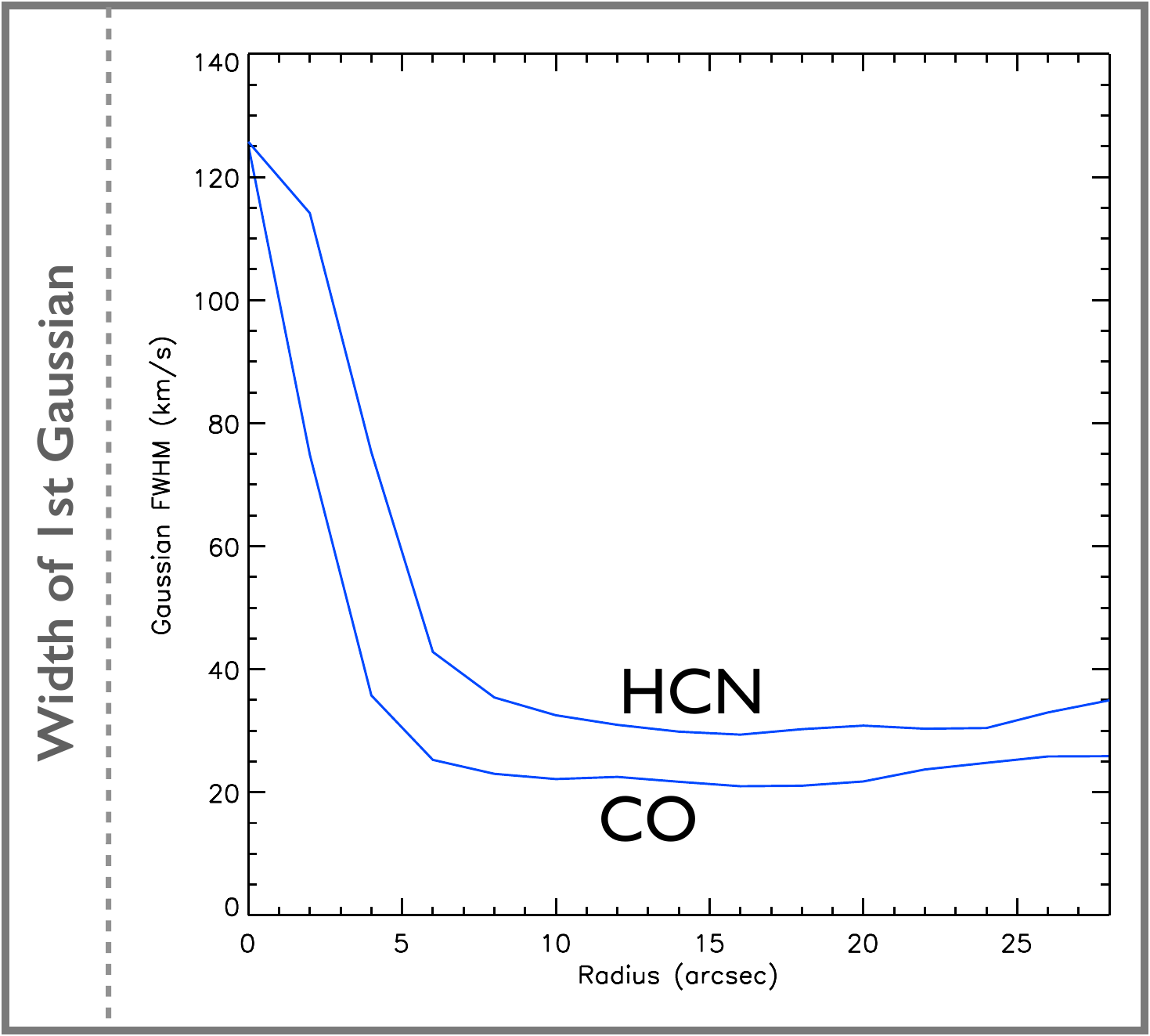}
\end{center}
\caption{Width of the first fitted Gaussian (the one that carries most flux, Sect.\,\ref{Sec:multiplecomp}) as a function of radius.
}
\label{fig:widthGauss1}
\end{figure}

\begin{figure*}[t]
\begin{center}
\includegraphics[width=1.0\textwidth]{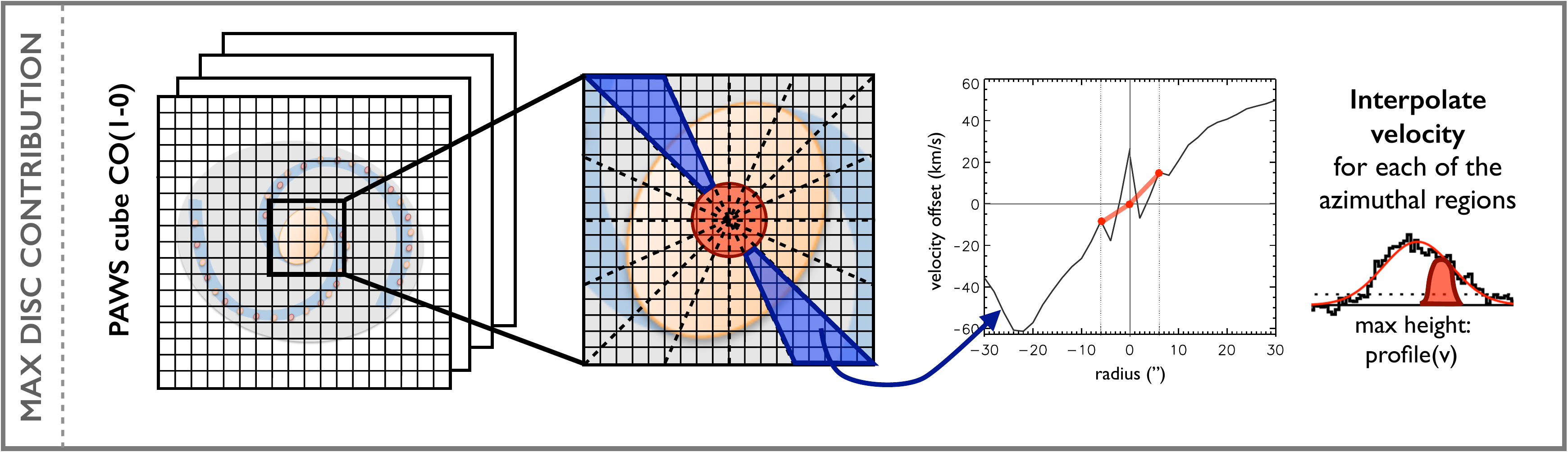}
\end{center}
\caption{In the central $r \lesssim 5''$ it becomes extremely hard to separate a line associated with the disc from the extremely broad outflowing component that dominates its flux.
This is a visual illustration of how we estimate the maximum emission from disc, assuming that its emission line has the same width as in the rest of the disc, $\sim$25\,km/s, and that its velocity centroid can be linearly interpolated in this small region from the surroundings. We perform this interpolation along 16 different azimuthal regions, forcing the central position to coincide with the systemic velocity, 472\,km/s.
}
\label{fig:DiscContr}
\end{figure*}

Here we estimate the maximum contribution from a regularly rotating \textit{disc} component to the molecular emission in the central $\sim$5$''$ of M51.
Indeed, as we move towards the centre of the galaxy, there is a point at which one can no longer distinguish the Gaussian line associated with the disc from the outflowing component, which becomes more and more dominant in flux (Sect.\,\ref{Sec:multiplecomp}). This is obvious from Fig.\,\ref{fig:widthGauss1}, where it becomes clear that the width of the first fitted Gaussian increases dramatically below $r \lesssim 5''$ ($r \lesssim 7''$ for HCN).
In this central region, we estimate the maximum possible contribution from the disc (as illustrated in Fig.\,\ref{fig:DiscContr}) following a simple argument. We assume that the disc component has approximately the same line width as outside that central region: as we have seen (Fig.\,\ref{fig:widthGauss1}), the line width at $r \gtrsim 5''$ is well confined between 20--30\,km/s for CO ($r \lesssim 7''$ for HCN, widths in the range 30--40\,km/s); therefore, we assume an intermediate characteristic width of 25\,km/s for the hypothetical disc contribution in this central area (35\,km/s for HCN). 
Then, we interpolate its velocity centroid linearly in this small region, along 16 different azimuthal regions, starting from the average value in the immediate vicinity (at $r = 5''$ or $r = 7''$), and imposing that  the central position must coincide with the systemic velocity of the galaxy ($v_\mathrm{sys}=472$\,km/s). We have confirmed that varying the number of azimuthal zones or extending the maximum radius out to which we interpolate does not affect the velocity centroids significantly ($\Delta v \lesssim 5$\,km/s). 
The peak of this Gaussian is given by the maximum height allowed by the actual line profile at the interpolated velocity centroid (Fig.\,\ref{fig:DiscContr}).
The flux of this interpolated Gaussian is assigned, as an upper limit, to the first Gaussian (``disc''), and the remaining flux, as a lower limit, to the second Gaussian (``outflow'').



\section{Outflow Geometry}
\label{angle}

The angle $\alpha$ that is part of the formula to calculate the outflow rate (Eq.\,\ref{eqn:outflow}) is well-constrained for M51. From careful modelling of the observed kinematics in the southern XNC, \citet{1988ApJ...329...38C} concluded that the shock working surface that they analyse is inclined $\sim$20$^\circ$ to the line of sight. This result is robust, in any case well-bounded between 20$^\circ$ and 25$^\circ$, and compatible with a number of additional constraints (disc scale-height, observed velocities, etc.). It is important to emphasise, though, that to our best knowledge there is no compelling reason to assume that the southern cone is near us, as we discuss  in Sect.\,\ref{Sec:distribution}, Sect.\,\ref{HCN-COdiff}.
In any case this implies that, if the axis of the jet is perpendicular to the shock working surface, as expected, the angle between the jet and the line of sight must be 70$^\circ$.

We are now in a position to deproject this value (and assess uncertainties) and provide the true value of the inclination of the jet with respect to the plane of the disc. The scalar product of a (unitary) vector pointing along the axis of the jet and a (unitary) vector normal to the plane of the galaxy is:

\begin{equation}
\mathbf{j} \cdot \mathbf{n} = 
 \begin{pmatrix}
  \cos{\alpha} \\
  \sin{\alpha} \cos{\phi} \\
  -\sin{\alpha} \sin{\phi}
 \end{pmatrix}
 \begin{pmatrix}
  \cos{i} \\
  0 \\
  \sin{i}
 \end{pmatrix}
 =\sin{\xi}
\end{equation}

For the geometry assumed in Fig.\,\ref{fig:geometry} (corresponding to the orientation of M51), where $\alpha$ is the (smallest) angle between the jet and the line of sight,  $\phi$ is the {\it projected} azimuthal angle between the jet axis and the line of nodes, and $i$ is the inclination of the galaxy with respect to the line of sight ($i=0$ for face-on), and $\xi$ is the (smallest) angle between the jet axis and the plane of the galaxy (from the definition of scalar product, $\mathbf{j} \cdot \mathbf{n} = |\mathbf{j}| \, |\mathbf{n}| \, \cos{\widehat{jn}}$, with $\widehat{jn} = 90^\circ - \xi$). In the case of M51, adopting the orientation of the jet measured by \citet{2004ApJ...603..463B} (PA$=163^\circ$, with a projected opening angle of 74$^\circ$), and the inclination of the galaxy determined by PAWS \citep{2014ApJ...784....4C} this expression reduces to:

\begin{equation}
\mathbf{j} \cdot \mathbf{n} = 
 \begin{pmatrix}
  \cos(70^\circ) \\
  \sin(70^\circ) \cos(10^\circ) \\
  -\sin(70^\circ) \sin(10^\circ)
 \end{pmatrix}
 \begin{pmatrix}
  \cos(22^\circ) \\
  0 \\
  \sin(22^\circ)
 \end{pmatrix}
  =\sin{\xi}
\end{equation}

Which implies an angle of 15$^\circ$ between the jet and the plane of the galaxy. As a sanity check, we can make use of the gaseous disc scaleheights measured by \citet{2013ApJ...779...43P} and the extent of the narrow-line emission clearly associated with the biconic jet \citep{2004ApJ...603..463B} to obtain an independent estimation of the angle between the outflow and the plane of the galaxy. We assume that the narrow-line emission comes from the interaction between radio jet and thin gaseous disc, roughly extending vertically up to the scale-height of the molecular gas traced by PdBI at that radius. In the maps from \citet{2004ApJ...603..463B}, both the radio emission and the NLR emission closely follow a (projected) biconic geometry, and drop off quite sharply at a (projected) radius of 4''. This corresponds to a (deprojected) radius of 148\,pc, and at this radius, the scale-height of the thin disc is $\sim$35\,pc. If the limit of the NLR and radio continuum correspond to this scale-height, this would imply an angle of 13$^\circ$, in good agreement with the value of 15$^\circ$ implied by \citet{1988ApJ...329...38C}.
This is also in agreement with the fact that M51 is a Seyfert 2 galaxy; thus, the inclination of the jet could not be much higher or we would, otherwise, see the broad line region.

\end{document}